\documentclass[prd,nofootinbib,preprint,superscriptaddress]{revtex4}
\pdfoutput=1
\usepackage[T1]{fontenc}
\usepackage{amsmath,amssymb}
\usepackage{epsfig}
\usepackage{subfigure}
\usepackage{float}
\usepackage{graphicx}
\usepackage[usenames,dvipsnames]{color}
\usepackage{slashed}
\usepackage{multirow}
\usepackage[colorlinks,citecolor=blue]{hyperref}
\usepackage{pdfpages}
\usepackage{color}
\usepackage{comment}

\newcommand{\mhsqr}[1]{M_{H_1}^2}
\newcommand{\boldA}{\bf A}
\begin{document}
\title{Light Dirac neutrino portal dark matter with observable $\Delta{N_{\rm eff}}$}
\author{Anirban Biswas}
\email{anirban.biswas.sinp@gmail.com}
\affiliation{Centre of Excellence in Theoretical
and Mathematical Sciences,
Siksha `O'Anusandhan (Deemed to be University),
Khandagiri Square, Bhubaneswar 751030,
Odisha, India}
\author{Debasish Borah}
\email{dborah@iitg.ac.in}
\affiliation{Department of Physics, Indian Institute of Technology
Guwahati, Assam 781039, India}
\author{Dibyendu Nanda}
\email{dibyendu.nanda@iitg.ac.in}
\affiliation{Department of Physics, Indian Institute of Technology
Guwahati, Assam 781039, India}
\begin{abstract}
We propose a Dirac neutrino portal dark matter scenario
by minimally extending the particle content of the Standard Model
(SM) with three right handed neutrinos ($\nu_R$), a Dirac fermion
dark matter candidate ($\psi$) and a complex scalar ($\phi$), all
of which are singlets under the SM gauge group. An additional $\mathbb{Z}_4$
symmetry has been introduced for the stability of dark matter
candidate $\psi$ and also ensuring the Dirac nature of light
neutrinos at the same time. Both the right handed neutrinos and
the dark matter thermalise with the SM plasma due to a new Yukawa
interaction involving $\nu_R$, $\psi$ and $\phi$ while the latter
maintains thermal contact via the Higgs portal interaction.
The decoupling of $\nu_R$ occurs when $\phi$
loses its kinetic equilibrium with the SM plasma and thereafter
all three $\mathbb{Z}_4$ charged particles form an equilibrium
among themselves with a temperature $T_{\nu_R}$. 
The dark matter candidate $\psi$ finally freezes out within the dark
sector and preserves its relic abundance. We have found that in the
present scenario, some portion of low mass dark matter ($M_{\psi}\lesssim10$ GeV)
is already excluded by the Planck 2018 data for keeping $\nu_R$s
in the thermal bath below a temperature of 600 MeV and thereby
producing an excess contribution to $N_{\rm eff}$. The next
generation experiments like CMB-S4, SPT-3G etc.
will have the required sensitivities to probe the entire
model parameter space of this minimal scenario,
especially the low mass range of $\psi$ where direct detection
experiments are still not capable enough for detection.
\end{abstract}
\maketitle
\section{Introduction}
\label{sec:Intro}
Evidences from astrophysics and cosmology based experiments
suggest the presence of a non-baryonic, non-luminous form of
matter in the universe comprising approximately 26\% of its
energy density \cite{Zyla:2020zbs, Aghanim:2018eyx}. In terms of density 
parameter $\Omega_{\rm DM}$ and $h = \text{Hubble Parameter}/(100 \;\text{km}
~\text{s}^{-1} \text{Mpc}^{-1})$, the present abundance of this
form of matter, popularly known as dark matter (DM),
is conventionally reported as \cite{Aghanim:2018eyx}:
$\Omega_{\text{DM}} h^2 = 0.120\pm 0.001$
at 68\% CL. Given that DM has a particle origin, it is known that none of the Standard Model (SM) particles can satisfy all the criteria of a particle DM candidate. This has led to several beyond standard model (BSM) proposals out of which the the weakly interacting massive particle (WIMP) paradigm is perhaps the most widely studied one. In this framework, a DM particle having masses and interactions similar to those around the electroweak scale gives rise to the observed relic after thermal freeze-out, a remarkable coincidence often referred to as the {\it WIMP Miracle} \cite{Kolb:1990vq}. For a review of WIMP type models, please see \cite{Arcadi:2017kky} and references therein. 

In addition to DM, the SM also can not explain the origin of neutrino mass and mixing, as verified at neutrino oscillation experiments \cite{Zyla:2020zbs, Mohapatra:2005wg}. In spite of such evidences suggesting tiny neutrino mass and large leptonic mixing \cite{Esteban:2018azc}, the nature of neutrino: Dirac or Majorana, is not yet known. While neutrino oscillation experiments can not settle this issue, there are other experiments like the ones looking for neutrinoless double beta decay ($0\nu \beta \beta$), a promising signature of Majorana neutrinos. However, there have been no such observations yet which can confirm Majorana nature of light neutrinos. This has led to growing interest in studying the possibility of light Dirac neutrinos even though the conventional neutrino mass models have focussed on Majorana neutrino scenarios for last several decades. Such BSM framework must be invoked to explain non-zero neutrino mass as in the SM, there is no way to couple the neutrinos to the Higgs field in the renormalisable Lagrangian due to the absence of right handed neutrinos. While conventional neutrino mass models based on seesaw mechanism can be found in \cite{Minkowski:1977sc, GellMann:1980vs, Mohapatra:1979ia, Schechter:1980gr, Mohapatra:1980yp, Lazarides:1980nt, Wetterich:1981bx, Schechter:1981cv, Brahmachari:1997cq, Foot:1988aq}, scenarios describing light Dirac neutrino mass may be found in \cite{Babu:1988yq, Peltoniemi:1992ss, Chulia:2016ngi, Aranda:2013gga, Chen:2015jta, Ma:2015mjd, Reig:2016ewy, Wang:2016lve, Wang:2017mcy, Wang:2006jy, Gabriel:2006ns, Davidson:2009ha, Davidson:2010sf, Bonilla:2016zef, Farzan:2012sa, Bonilla:2016diq, Ma:2016mwh, Ma:2017kgb, Borah:2016lrl, Borah:2016zbd, Borah:2016hqn, Borah:2017leo, CentellesChulia:2017koy, Bonilla:2017ekt, Memenga:2013vc, Borah:2017dmk, CentellesChulia:2018gwr, CentellesChulia:2018bkz, Han:2018zcn, Borah:2018gjk, Borah:2018nvu, CentellesChulia:2019xky,Jana:2019mgj, Borah:2019bdi, Dasgupta:2019rmf, Correia:2019vbn, Ma:2019byo, Ma:2019iwj, Baek:2019wdn, Saad:2019bqf, Jana:2019mez, Nanda:2019nqy} and references therein.

Thus, in order to realise light Dirac neutrinos at sub-eV scale
as well as DM in the universe we need to extend the SM at least by
two different types of fields: three singlet right chiral neutrinos
and the DM field. The inclusion of such ultra-light degrees
of freedom (DOF) has a deep impact on the cosmological evolution
of the universe as such DOF, depending on their era of decoupling
from the SM bath, can contribute immensely to the radiation energy
density ($\varrho_{\rm rad}$). This results in an alteration in
the expansion rate of the universe since the Hubble parameter
during the radiation dominated era is $\mathcal{H} \simeq 
\sqrt{\dfrac{8 \pi}{3\,m^2_{\rm pl}}\,\varrho_{\rm rad}}$, where
$m_{\rm pl} = 1.22\times 10^{19}$ GeV is the Planck mass. This will
further lead to observable signatures through modifications in
the primordial abundances of light elements such as
$^4$He, D and $^7$Li as predicted by the Big Bang Nucleosynthesis (BBN)
and also deformation in the Cosmic Microwave Background Radiation (CMB)
power spectrum during the era of recombination. Consequently, there is
no room for new physics that introduces fully thermalised additional
relativistic species which remain in thermal contact with the SM
at the onset of nucleosynthesis ($T\lesssim\mathcal{O}({\rm MeV})$).
However, in addition to the SM particles, extra relativistic species 
decoupled at $T\gtrsim \mathcal{O}(100\,{\rm MeV})$ are
still allowed by the current data on $N_{\rm eff}$ from the Planck
satellite \cite{Aghanim:2018eyx}, where $N_{\rm eff}$ is the effective
number of relativistic species (except photon) contributing to
the radiation energy density. The quantity $N_{\rm eff}$ is defined
as the contribution of non-photon components to the radiation energy
density normalised by the contribution of a single active neutrino
species ($\varrho_{\nu_L}$) \cite{Mangano:2005cc}
i.e.
\begin{eqnarray}
N_{\rm eff} &\equiv& \dfrac{\varrho_{\rm rad}-\varrho_{\gamma}}
{\varrho_{\nu_L}}\,.
\label{def_neff}
\end{eqnarray}
Recent 2018 data from the CMB measurement by the Planck
satellite \cite{Aghanim:2018eyx} suggest that the effective
degrees of freedom for neutrinos during the
era of recombination ($z\sim 1100$) as 
\begin{eqnarray}
{\rm
N_{eff}= 2.99^{+0.34}_{-0.33}
}
\label{Neff}
\end{eqnarray}
at $2\sigma$ or $95\%$ CL including baryon acoustic oscillation (BAO) data.
At $1\sigma$ CL it becomes more stringent to $N_{\rm eff} = 2.99 \pm 0.17$.
Both these bounds are consistent with the standard model (SM)
prediction\footnote{The deviation from $N_{\rm eff}=3$
is due to various effects like non-instantaneous
neutrino decoupling, flavour oscillations
and finite temperature QED corrections to
the electromagnetic plasma \cite{Froustey:2020mcq, Mangano:2005cc, Mangano:2001iu}.}
$N^{\rm SM}_{\rm eff}=3.045$ \cite{Mangano:2005cc, Grohs:2015tfy,
deSalas:2016ztq}. Upcoming CMB Stage IV (CMB-S4)
experiments are expected to put much more stringent
bounds than the Planck experiment due
to their potential of probing all the way down to
$\Delta N_{\rm eff}=N_{\rm eff}-N^{\rm SM}_{\rm eff}
= 0.06$ \cite{Abazajian:2019eic}.

Nevertheless, there are still some room for the physics
beyond the SM and we are exploring one of the well motivated
possibilities where dark matter thermalises with
the SM bath through right handed neutrinos and vice-versa. Our
dark matter candidate is a Dirac fermion and it belongs to
the class of WIMP dark matter. Typical WIMP type DM models have different portals
via which DM can interact with the SM bath. In our work,
DM couples to the SM only via light Dirac neutrinos
and we call it Dirac neutrino portal dark matter (DNPDM). Apart from
this minimal setup connecting light Dirac neutrino and DM simultaneously,
there exist two other motivation for such scenario. Firstly, since DM
couples to SM only via light Dirac neutrinos or the right chiral parts
of Dirac neutrinos to be more specific, there is no tree level DM-nucleon
coupling keeping the model safe from stringent direct detection
bounds \cite{Aprile:2017iyp, Aprile:2018dbl}. Secondly, thermalisation of DM
will also lead to thermalisation of right handed neutrinos giving rise
to additional contribution to the relativistic degrees of freedom in
the early universe. 
For some recent studies on light Dirac neutrinos and enhanced $\Delta N_{\rm eff}$ in different contexts, please see \cite{Abazajian:2019oqj, FileviezPerez:2019cyn, Nanda:2019nqy, Han:2020oet, Luo:2020sho, Borah:2020boy, Adshead:2020ekg, Luo:2020fdt, Mahanta:2021plx, Du:2021idh}. It should also be noted that neutrino portal DM have been studied in different contexts by several authors, for example, see \cite{Falkowski:2009yz, Macias:2015cna, Batell:2017rol, Batell:2017cmf, Bandyopadhyay:2018qcv, Chianese:2018dsz, Blennow:2019fhy, Lamprea:2019qet, Chianese:2019epo, Bandyopadhyay:2020qpn, Hall:2019rld, Berlin:2018ztp}. However, in these works either DM coupling directly with the SM lepton doublet was considered or a portal via heavy right handed or Dirac neutrinos was discussed. While these scenarios can have interesting signatures, specially for indirect detection experiments, they are different from our proposal in this work both from the model as well as the phenomenology point of view.

We first consider a minimal model where BSM fields are limited to
three right handed neutrinos, one additional singlet fermion
(which is our DM candidate) and one additional singlet scalar
to facilitate the coupling of DM with right handed neutrinos. Additional
discrete symmetry $\mathbb{Z}_4$ is imposed to allow desired couplings
of these fields within themselves as well as with the SM particles. Focusing
on the WIMP type scenario, we then find the parameter space leading to
correct DM relic abundance and then calculate the contribution to $\Delta N_{\rm eff}$
for the same set of parameters. We show how a part of the parameter space
consistent with relic density requirement is already ruled out by Planck 2018
bounds on $\Delta N_{\rm eff}$ while the remaining parameter space remains
completely within the reach of future CMB experiments. While in the minimal
model light Dirac neutrino mass arises from the SM Higgs field only
with fine-tuned Yukawa couplings, we also briefly comment on the
possibility of a neutrinophilic Higgs doublet with induced vacuum
expectation value (VEV) towards the end where similar phenomenology
can be realised with less severe fine-tuning.

This paper is organised as follows. In Section\,\,\ref{sec:model},
we briefly discuss our minimal model of Dirac neutrino portal
dark matter. The Section\,\,\ref{sec:DM&RD} is devoted to
the discussions on our dark matter candidate
and necessary Boltzmann equations required for computing
relic density and dark sector sector. A detail discussion
on $\Delta{N}_{\rm eff}$ has been presented in
Section\,\,\ref{sec:delNeff}. Our numerical results are 
given in Section\,\,\ref{sec:res}. Finally, we conclude
in Section\,\,\ref{sec:conclu}. A detail derivation of
the Boltzmann equation expressing evolution of
dark sector temperature and expressions of necessary
annihilation cross sections are given in Appendices\,\,\ref{App:BE}
and \ref{App:2to2xsections}.
\section{The Model}
\label{sec:model}
In this section, we briefly discuss our model, the relevant particle
spectrum and interactions. As mentioned above, our primary motivation
is to constrain the DM parameter space from the Dirac nature of neutrinos.
Keeping that in mind, we have introduced one new Yukawa interaction
involving three new fields, two fermionic fields $\nu_R,\,\,\psi$
and one singlet complex scalar $\phi$. While $\nu_R$ is a right
chiral field, the other fermion singlet $\psi$ is considered to
be a Dirac fermion. All these new fields transform as singlets
under the SM gauge symmetry. Additionally, we have imposed
a discrete symmetry $\mathbb{Z}_4$ with the SM gauge symmetries
and the assigned charges are given in the table \ref{tab:1}.
Moreover, we have set $\mathbb{Z}_4$ charge $i$ to all the SM leptons to ensure
only the Dirac mass of neutrinos through the Higgs mechanism, while forbidding
the Majorana mass term of right handed neutrinos. The SM fields
not included in table \ref{tab:1} transform trivially under
$\mathbb{Z}_4$. Furthermore, the $\mathbb{Z}_4$ symmetry remains unbroken
ensuring the stability of DM. Although here we stick to this minimal setup,
we need at least one additional singlet scalar field ($\phi$) so that a
renormalisable interaction between DM ($\psi$) and right handed
neutrinos ($\nu_R$) can be achieved. 
\begin{table}
\begin{center}
\begin{tabular}{|c|c|c|c|}
\hline
Particles & $SU(3)_c \times SU(2)_L \times U(1)_Y$ &\ \  $\mathbb{Z}_4$\\
\hline
$\ell^{\alpha}_L$ & $(1, 2, -\frac{1}{2})$ & $i$\\
\hline
$e^{\alpha}_R$ & $(1, 1, -1)$ & $i$\\
\hline
$\nu^\alpha_{R}$ & $(1, 1, 0)$ & $i$\\
\hline
$\psi$ & $(1, 1, 0)$ &  $-1$\\
\hline
$\phi$ & $(1, 1, 0)$ & $i$\\
\hline
\end{tabular}
\end{center}
\caption{Fermion and scalar fields of the model charged
under $\mathbb{Z}_4$ symmetry.}
\label{tab:1}
\end{table}

The Lagrangian for the non-SM fields of this model, invariant under
the full symmetry group, is given as 
\begin{eqnarray}
\mathcal{L} \supset \mathcal{L}_{fermion} + \mathcal{L}_{scalar} \,.
\label{model:Lag:total}
\end{eqnarray}
The first term $\mathcal{L}_{fermion}$ is the Lagrangian for the new
fermion fields including the interaction with the SM leptons. The
expression of $\mathcal{L}_{fermion}$ is given by,
\begin{eqnarray}
\mathcal{L}_{fermion} = i \,\overline{\nu}_R\, \gamma^\mu \, \partial_\mu \, \nu_R\,
+\,  i \,\overline{\psi}\, \gamma^\mu \, \partial_\mu \, \psi\,
- \, M_{\psi} \overline{\psi} \psi - \left(y_H \, \overline{\ell}\, \tilde{H}\, \nu_R 
+  y_\phi \,\overline{\psi}\,{\nu}_R\, \phi + {\rm h.c.} \right)\,.
\end{eqnarray}
The first two terms are the kinetic terms for $\nu_R$ and $\psi$ respectively
while the third term is the bare mass term for Dirac fermion $\psi$ which
is playing the role of dark matter in this model. As discussed above, the
light neutrino mass arises through the conventional Higgs mechanism from
the interaction $y_H \, \overline{\ell}\, \tilde{H}\, \nu_R$.
The required Yukawa coupling for generating sub-eV scale neutrino mass is of
the order of $10^{-12}$ \footnote{As noted above, there are several
UV complete realisations of Dirac neutrino models where such fine
tunings can be avoided at the cost of incorporating more fields. Here we
stick to the minimal setup for simplicity.}. Finally, the last term represents the most important interaction relevant to the phenomenology we discuss here.
All three non-SM fields are interacting among themselves via the last term
in the above Lagrangian. 
 
The second term in Eq.\,(\ref{model:Lag:total}) contains the gauge
invariant interactions between the scalar fields (including the Lagrangian for
the SM Higgs doublet $H$) as given by 
\begin{eqnarray}\nonumber
\mathcal{L}_{scalar} &=& (D_{H\mu} H)^\dagger (D_{H}^\mu H)
+ (\partial_{\mu} \phi)^\dagger (\partial^\mu \phi) - 
\Bigg[ - {\mu_{H}^2}\,( H^\dagger H)  + {\lambda_{H}}\, 
( H^\dagger H)^2 + {\mu_{\phi}^2}\,( \phi^\dagger \phi) + \\ &&
\lambda_{\phi} \, ( \phi^\dagger \phi)^2 + \lambda_{H \phi}\,
(H^\dagger H)(\phi^\dagger \phi) + \lambda_{\phi}^\prime 
\left(\phi^4 + (\phi^\dagger)^4\right)\Bigg]\,,
\label{potential}
\label{scalar:pot}
\end{eqnarray}

where, the covariant derivative for $H$ is defined as 
\begin{eqnarray}
D_{H\mu}H & = & \left(\partial_\mu + i \frac{g}{2}\sigma_a W^a_{\mu} 
+ i \frac{g^\prime}{2} B_{\mu} \right) H\,.
\end{eqnarray}
Here, $g$ and $g^\prime$ are the gauge couplings
for $SU(2)_{L}$ and $U(1)_Y$ respectively while the
corresponding gauge bosons are denoted by $W_{\mu}^a$ and $B_{\mu}$.
The complex scalar singlet $\phi$ does not acquire any vacuum expectation
value. However, as in the SM, the neutral component
of the Higgs doublet ${H}$ acquires a non-zero VEV
with $v=246$ GeV and $SU(2)_L \times U(1)_Y$ symmetry is
spontaneously broken. The representation of $H$ in
the unitary gauge is given by,
\begin{eqnarray}
H=\begin{pmatrix}0\\
\dfrac{h + v}{\sqrt{2}}\end{pmatrix}\,.\,
\label{H&phi_broken_phsae}
\end{eqnarray}
Minimization condition of the above potential will come out as the following,
\begin{eqnarray}
-\mu_{H}^2 + 2 \lambda_{H} v^2 = 0\,.
\end{eqnarray}
By using the above condition, the masses of the physical
scalars can be written as,
\begin{eqnarray}
M_{h}^2 &=& 2 \lambda _{H}\, v^2\,, \\
M_{\phi}^2 &=& \mu_{\phi}^2 + \frac{1}{2} v^2 \lambda_{H\phi}^2\,.
\end{eqnarray}
The free parameters of this model are the following couplings and the masses,
\begin{equation}
{\rm M_{\phi}\,,M_{\psi}\,,y_\phi, \lambda_{H\phi},\lambda_\phi\,,
\text{and}\,\, \lambda_\phi^\prime}
\end{equation}

The portal coupling $\lambda_{H\phi}$ has utmost importance here
as it is the sole connector between the SM sector and the dark sector. While $\nu_R$ also acts like a portal between these sectors, the scalar portal coupling $\lambda_{H\phi}$ directly connects them. The complex scalar $\phi$ thermalises with the SM bath through elastic
scatterings like $\phi X \rightarrow \phi X$, with $X$ being
the SM fermions, gauge bosons and the Higgs boson.
All these interactions involve the portal coupling $\lambda_{H\phi}$.
In the present model, other dark sector field $\psi$ do not have any direct coupling with
the SM fields. While $\nu_R$ has direct coupling with SM leptons via the Higgs, the corresponding Yukawa couplings are too tiny (in order to satisfy neutrino mass constraints) to bring this interaction into equilibrium. In spite of that, they can still maintain thermal
equilibrium with the SM bath by virtue of the new Yukawa
interaction involving $\phi$. Therefore, once $\phi$ thermalises
through the portal interactions, both $\nu_R$ and $\psi$
also share a common temperature with the SM bath.
The dark sector temperature will deviate from the photon
temperature when the kinetic equilibrium between
$\phi$ and the SM bath is lost. Thus, the presence of
$\phi$ in the thermal bath plays an important role
in the production of both DM ($\psi$) and $\nu_R$ as sizeable
value of the new Yukawa coupling $y_{\phi}$ will make sure thermal
production of both these species in the SM plasma. The presence of
extra light DOFs in the thermal plasma
during BBN and recombination is tightly constrained from the measurement
of cosmological parameter $\Delta N_{\rm eff}$ \cite{Aghanim:2018eyx}
which suggests that these new DOFs have to be decoupled from the SM plasma
much earlier than the left-handed neutrinos. Therefore,
in our analysis the portal coupling ($\lambda_{H\phi}$)
and the Yukawa coupling ($y_{\phi}$) will have significant
impact on the observables like $\Omega_{\psi} h^2$ and
$\Delta{N_{\rm eff}}$. We have a detailed discussion
about the thermalisation of the right handed neutrinos
and their contribution to $N_{\rm eff}$
in Section \ref{sec:delNeff}.
\section{Dark matter candidate and its relic density}
\label{sec:DM&RD}
In the present scenario,\,we have two $\mathbb{Z}_4$ charged fields
$\phi$ and $\psi$ other than the SM fields and three
$\nu_R$s.\,\,Since the $\mathbb{Z}_4$ symmetry remains
unbroken, therefore depending on the mass hierarchy, either $\phi$
or $\psi$ is absolutely stable and hence can be a possible
dark matter candidate. Here, we choose the Dirac fermion $\psi$ as
our dark matter candidate by considering $M_{\psi} < M_{\phi} + m_{\nu}$,
such that the only decay mode of $\psi$ into $\phi$ and $\nu_R$ becomes
kinematically forbidden. To check the viability of $\psi$ as
a dark matter candidate, first we need to compute its relic abundance at
the present epoch. However, the computation of relic density will
not be as straightforward as in the case of a normal WIMP
dark matter \cite{Gondolo:1990dk} since after the kinetic decoupling of $\phi$
from the SM bath the temperature of the dark sector is different from the photon temperature ($T$). Therefore, for $T<T_{\rm dec}$
we need to solve two coupled Boltzmann equations, one is for
the comoving number density $Y$ and the rest is for the
dark sector temperature $T_{\nu_R}$. As the new Yukawa interaction
between $\phi$, $\psi$ and $\nu_R$ is sufficiently strong
($y_{\phi} \sim \mathcal{O}(0.1)$), this helps to maintain a
thermal equilibrium among the three $\mathbb{Z}_4$ charged species
having a common temperature $T_{\nu_R}$. Here we have denoted
the dark sector temperature by the temperature of the
relativistic species $\nu_R$ similar to the SM where
the bath temperature is same as the photon temperature.
 
\begin{figure}[h!]
\includegraphics[height=6cm,width=7.5cm]{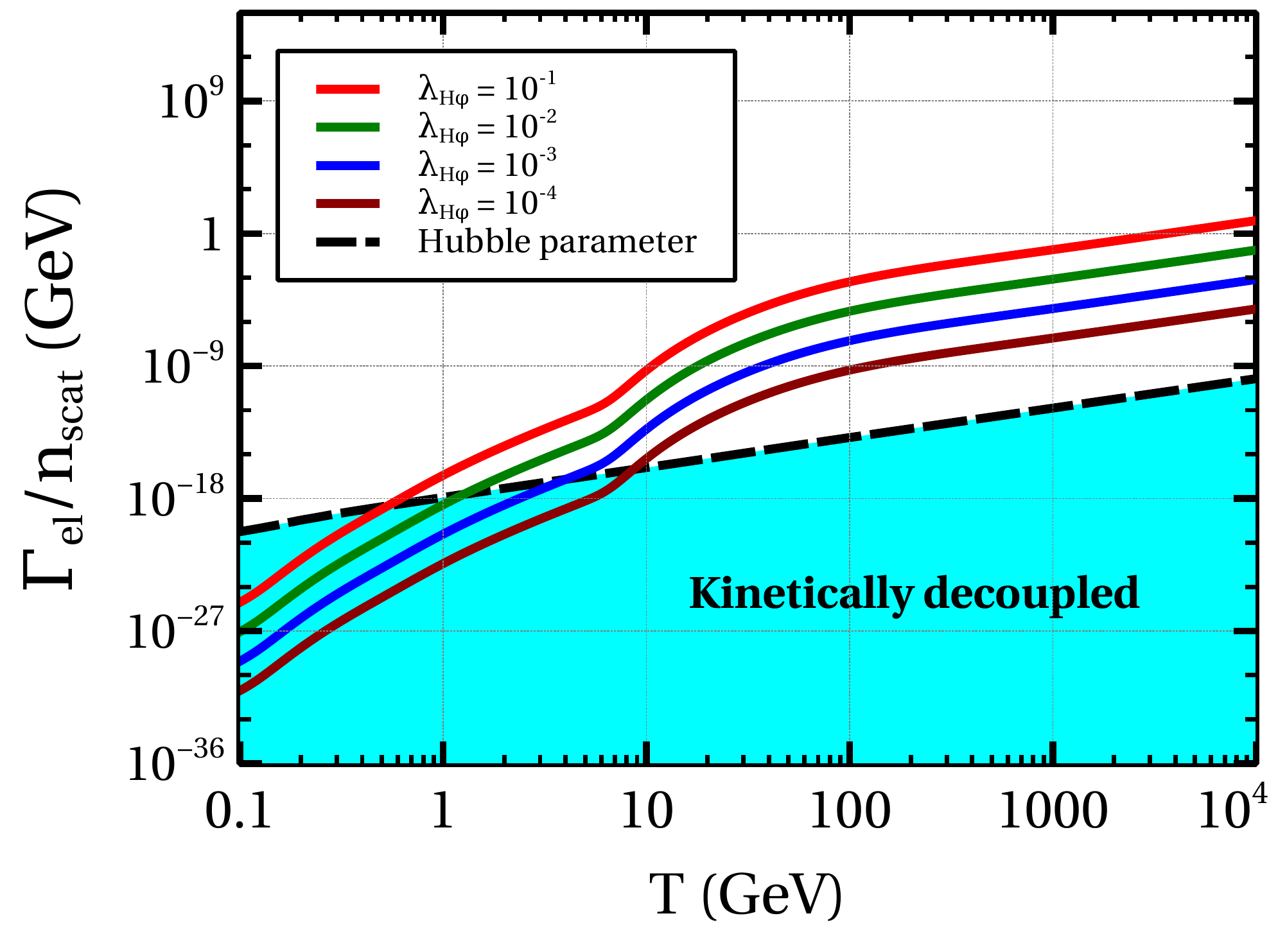}
\,\,\,\,\,\,\,\,
\includegraphics[height=6cm,width=7.5cm]{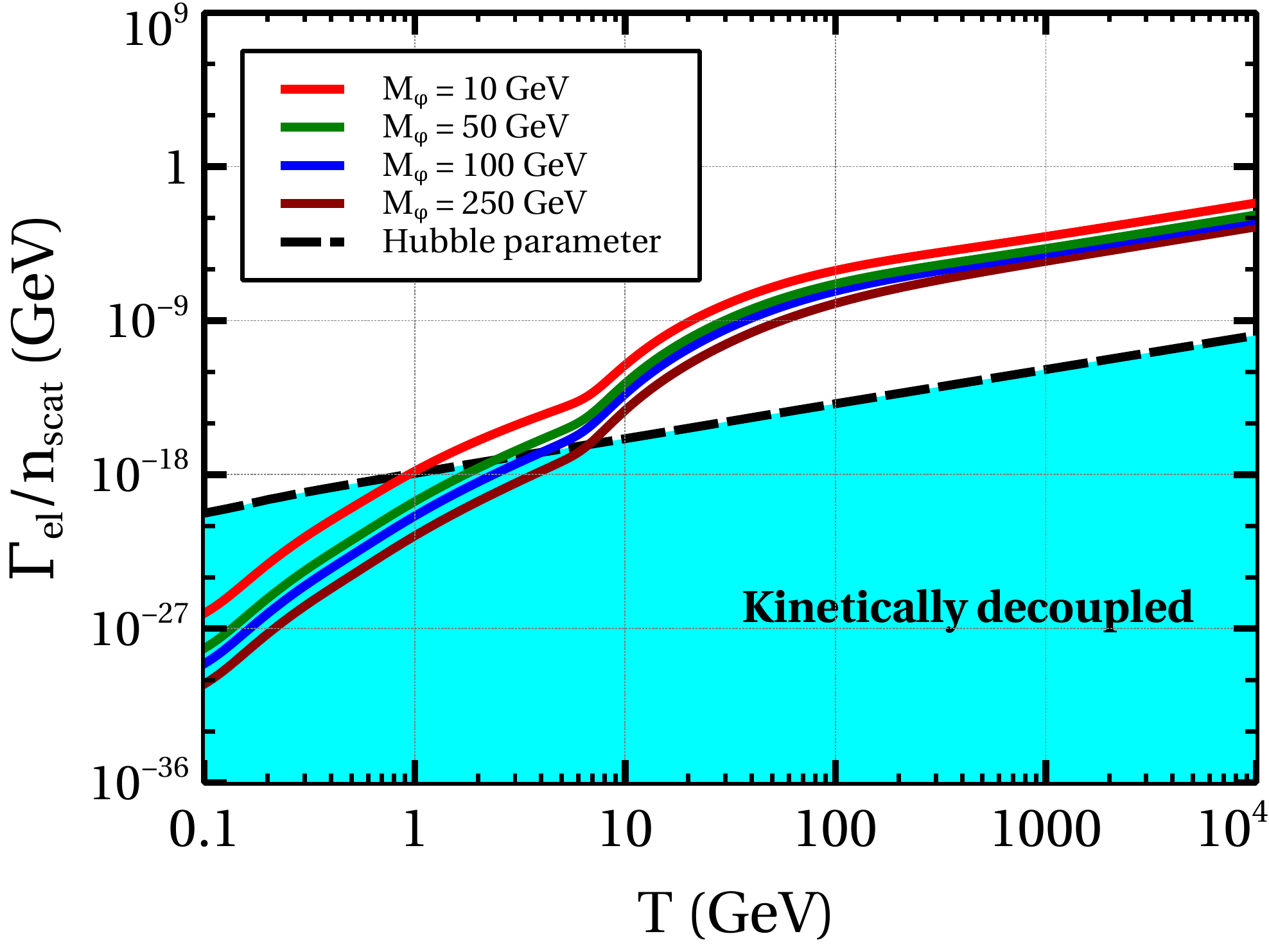}
\caption{Variation of the elastic scattering rate ($\Gamma_{\rm el}$)
normalised by $n_{\rm scat}$ with $T$ for different values of
$\lambda_{H\phi}$ (left panel) and $M_{\phi}$ (right panel) respectively.
Additionally, we have also shown the corresponding values of
$\mathcal{H}(T)$ by the black dashed line for comparison.}
\label{Fig:scattering_rate}
\end{figure}
\begin{figure}[h!]
\centering
\subfigure[Scatterings responsible for thermalisation
of $\nu_R$ within the dark sector.]
	{
      \includegraphics[height=3.1cm,width=14cm]
      {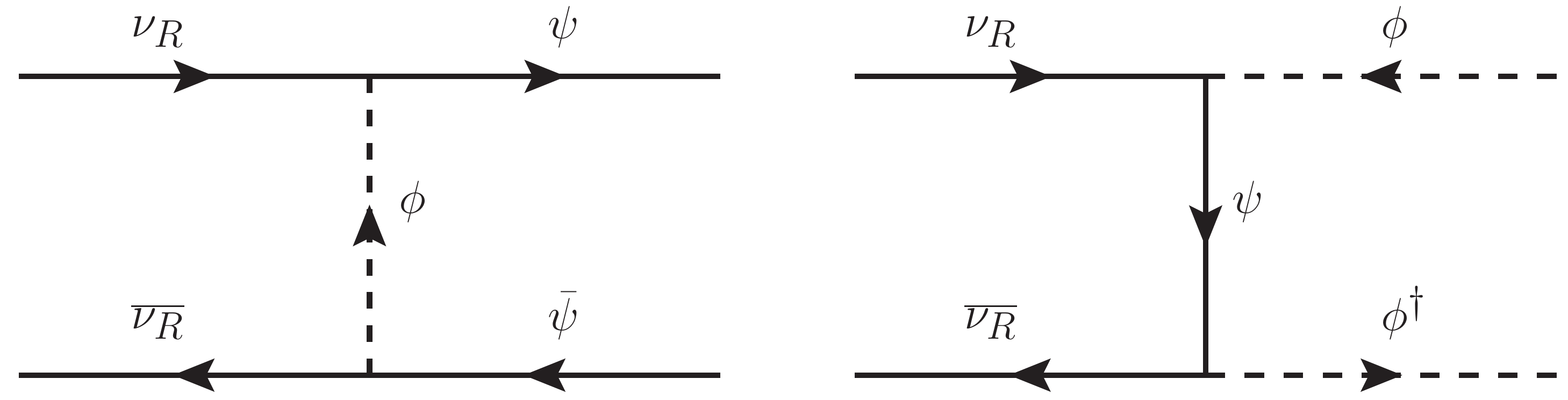}
     }
\vskip 0.2cm
\subfigure[Thermalisation processes of $\phi$ with the SM bath.]
	{
      \includegraphics[height=3.5cm,width=14cm]
      {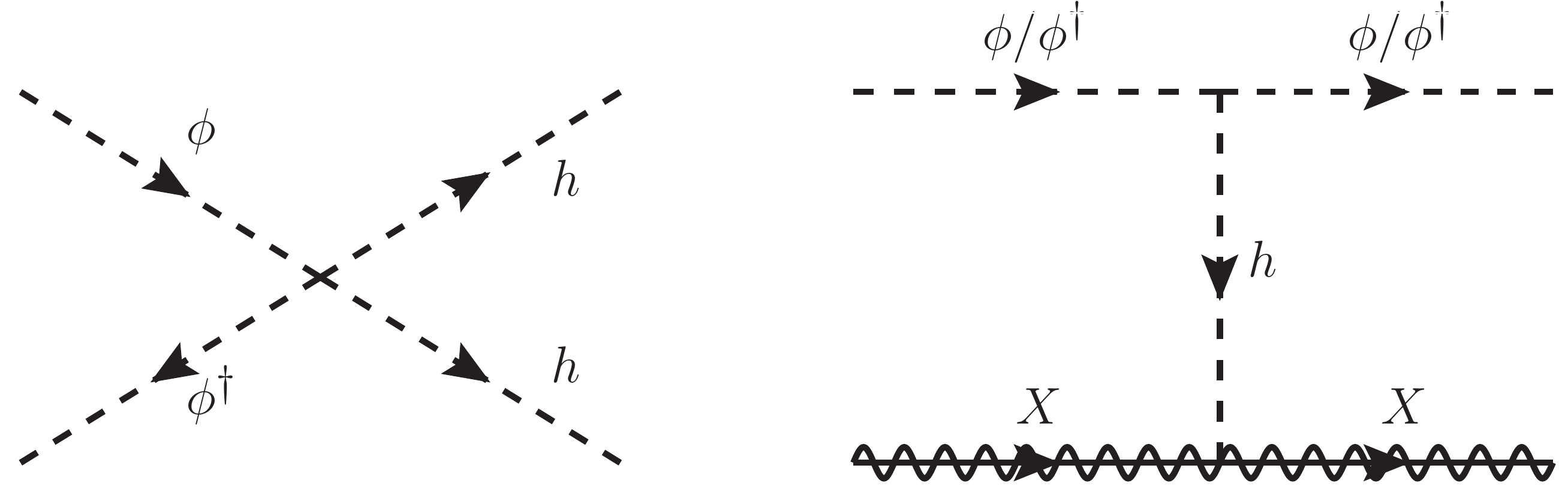}
    }
\caption{Feynman diagrams for the relevant scatterings.
Here $X$ is any SM particles (fermion or boson).}
\label{Fig:feyn_dia1}
\end{figure}

The decoupling temperature $T_{\rm dec}$, beyond which $\phi$
can no longer be in kinetic equilibrium with the SM bath, is obtained
approximately from $\dfrac{1}{n_{\rm scat}}\dfrac{\Gamma_{\rm el}}
{\mathcal{H}}\bigg|_{\scriptsize T_{\rm dec}}\simeq 1$
\cite{Gorbunov:2011zz, Gondolo:2012vh}. Here,
$\Gamma_{\rm el} = \sum_Xn^{\rm eq}_X\times\langle\sigma {\rm v}
\rangle_{\phi X \rightarrow \phi X}$ is the elastic
scattering ($\phi X \rightarrow \phi X$)
rate while $n^{\rm eq}_X$ being the equilibrium
number density of the SM species $X$. The Hubble parameter
controlling the expansion of the universe is denoted
by $\mathcal{H}$. Moreover, the quantity $n_{\rm scat}\sim\frac{M_{\phi}}{T}$
represents the number of scatterings required so that the
energy transfer between the SM bath and $\phi$ is
$\sim T$ \cite{Gorbunov:2011zz}. In Fig.\,\,\ref{Fig:scattering_rate},
we have depicted how the quantity $\Gamma_{\rm el}/n_{\rm scat}$
varies with $T$ for four different values of the portal
coupling $\lambda_{H\phi}$ and $M_{\phi}$ respectively. The expressions
of necessary scattering cross sections are given in Appendix\,\,\ref{App:2to2scatt}
and the corresponding Feynman diagrams are shown in the lower panel
of Fig.\,\,\ref{Fig:feyn_dia1}. Moreover, to understand the
era of kinetic decoupling of $\phi$ we have also shown
the variation of the Hubble parameter $\mathcal{H}(T)$
for the entire considered range $100\,\,{\rm MeV}\leq T\leq 10^4$ GeV.
From the left panel of Fig.\,\,\ref{Fig:scattering_rate}, we can
see that the decoupling of $\phi$ occurs between
$600\,\,{\rm MeV}\leq T_{\rm dec}\leq 10$ GeV for $M_{\phi}=100$ GeV
and $0.1\geq\lambda_{H\phi}\geq10^{-4}$. Similarly,
the mass of $\phi$ also plays a crucial role in determining
the decoupling temperature and that has been illustrated in
the right panel where we have kept $\lambda_{H\phi}$ fixed at
$10^{-3}$. In this case, for the increment of $M_{\phi}$ from 10 GeV
to 250 GeV, the corresponding $T_{\rm dec}$ changes
from 900 MeV to 4 GeV. 

Therefore, we have two regimes separated by the decoupling
temperature $T_{\rm dec}$. When $T\geq T_{\rm dec}$,
all the dark sector species maintain kinetic equilibrium
with the SM bath. Considering
$\frac{Y_{i}}{Y} \simeq \frac{Y^{eq}_{i}}{Y^{eq}}$ ($i = \phi,\,\,\psi$)
\cite{Griest:1990kh, Edsjo:1997bg}, we can reduce two coupled Boltzmann
equations for $\phi$ and $\psi$ into a single equation
involving the total comoving number density $Y = Y_{\phi} + Y_{\psi}$ i.e.
\begin{eqnarray}
\dfrac{dY}{dx} = - \dfrac{1}{2} \dfrac{\beta \,{\rm s}}{\mathcal{H}\,x}
\langle{\sigma {\rm v}} \rangle_{eff} \left[Y^2 - (Y^{eq})^2\right]\,,
\label{BEy_T_gt_Tdec}
\end{eqnarray}       
where, $x= \frac{M_0}{T}$ with $M_0$
being any arbitrary mass scale. Moreover, ${\rm s}$ is the entropy
density of the universe and $\beta(T) = \dfrac{g^{1/2}_{\star}(T)
\sqrt{g_{\rho}(T)}}{g_s(T)}$ with $g_s$ and $g_{\rho}$
being the effective DOFs associated with entropy and energy
densities respectively while $g^{1/2}_{\star} =
\dfrac{g_s}{\sqrt{g_{\rho}}}\left(1+\dfrac{1}{3}
\dfrac{T}{g_s}\dfrac{dg_s}{dT}\right)$.
Furthermore, the effective annihilation cross section
$\langle{\sigma {\rm v}} \rangle_{eff}$ is given by
\begin{eqnarray}
\langle{\sigma {\rm v}} \rangle_{eff} = \dfrac{
\langle{\sigma {\rm v}} \rangle_{\phi\phi^\dagger
\rightarrow X\overline{X},\,{\nu}_R \nu_R}\,
(Y^{\rm eq}_{\phi})^2 + 
\langle{\sigma {\rm v}} \rangle_{\psi\bar{\psi}\rightarrow {\nu}_R \nu_R}\,
(Y^{\rm eq}_{\psi})^2}
{\left(Y^{\rm eq}_{\phi} + Y^{\rm eq}_{\psi}\right)^2}\,.
\label{sigmaVeff}
\end{eqnarray}
Here, $\phi\phi^\dagger \rightarrow X\overline{X}$
\footnote{Similar
to the $\phi$ and $\phi^\dagger$ annihilations into the SM
particles, our dark matter candidate $\psi$ can also be pair
annihilated into the SM species. However, in this case
the Higgs portal coupling (see trilinear vertex
$\psi\overline{\psi}h$ in Eq.\,\,(\ref{gpsipsih}))
appears only at one loop level, making 
$\sigma_{\psi \overline{\psi}\rightarrow X\overline{X}}<<
\sigma_{\psi \overline{\psi}\rightarrow \nu_R\overline{\nu_R}}$
even at the resonance region ($M_{\psi}=M_h/2$).} represents
the pair annihilations of $\phi$ and $\phi^\dagger$ into
the SM species through the Higgs portal interaction and
the expressions of such cross sections for a real scalar field 
are given in \cite{Guo:2010hq, Biswas:2013nn}.

On the other hand, for $T<T_{dec}$, although the dark sector
has decoupled from the SM bath, they still maintains a local thermodynamic
equilibrium among $\phi$, $\psi$ and $\nu_R$ with a common
temperature $T_{\nu_R}\,(\neq T)$ due to sufficiently strong
Yukawa coupling $y_{\phi}$. The kinetic equilibrium within the
dark sector sustains well beyond the era of freeze-out by virtue of
adequate elastic scatterings $\nu_R + \phi(\psi)\rightarrow \nu_R+\phi(\psi)$.
However, besides $Y$, in this regime we need to solve an additional Boltzmann
equation for the dark sector temperature $T_{\nu_R}$.
Before writing these two Boltzmann equations, let us define a quantity
$\xi = \dfrac{T_{\nu_R}}{T}$. In terms of three dimensionless
quantities namely $Y$, $x$ and $\xi$, the Boltzmann equations
for $Y$ and $\xi$ are given by
\begin{eqnarray}
\dfrac{dY}{dx} &=& - \dfrac{1}{2} \dfrac{\beta \,{\rm s}}{\mathcal{H}\,x}
\langle{\sigma {\rm v}} \rangle_{eff} \left[Y^2 - (Y^{eq})^2\right]\,,
\label{BEy_T_lt_Tdec}\\
\hspace{-0.5cm}
x\,\dfrac{d\xi}{dx} + (\beta-1)\xi &=&
\dfrac{1}{2}\,\dfrac{\beta\,x^4\,{\rm s}^2}
{4\,\alpha\,\xi^3\,\mathcal{H}\,M_{0}^4}
\langle {E\sigma {\rm v}} \rangle_{eff}
\left[Y^2 - (Y^{eq})^2\right],
\label{BExi} 
\end{eqnarray} 
where, 
\begin{eqnarray}
\langle {E\sigma {\rm v}} \rangle_{eff} = 
\dfrac{(Y^{\rm eq}_{\psi})^2
\langle{E\,\sigma{\rm v}}\rangle^\prime_{
\nu_R \overline{\nu_R}\rightarrow\psi\overline{\psi}}\,
+ 
(Y^{\rm eq}_{\phi})^2
\langle{E\,\sigma{\rm v}}\rangle^\prime_{\nu_R \overline{\nu_R}
\rightarrow \phi\phi^\dagger}}
{\left(Y^{eq}_{\phi} + Y^{eq}_{\psi}\right)^2}\,
\label{esigmaVeff}
\end{eqnarray}
and $\alpha = g_{\nu_R}\times\dfrac{7}{8}\dfrac{\,\,\pi^2}{30}$
with $g_{\nu_R} = 2$. The difference between the two Boltzmann
equations (Eqs. (\ref{BEy_T_gt_Tdec}) and (\ref{BEy_T_lt_Tdec}))
is that for $T\geq T_{\rm dec}$, $Y$, $Y^{eq}$ and
$\langle {\sigma {\rm v}}\rangle_{eff}$ all are functions of $T$ (or $x$)
only while in the later case when $T$ drops below $T_{\rm dec}$
all these quantities depend on both $x$ and $\xi$. The evolution
of $\xi$ with respect to $x$ (equivalent to $T_{\nu_R}$ vs $T$)
is described by Eq.\,(\ref{BExi}) where similar to
$\langle {\sigma {\rm v}}\rangle_{eff}$, the quantity
$\langle {E\sigma {\rm v}}\rangle_{eff}$ also depends
on both $x$ and $\xi$. A detail derivation of the Boltzmann
equation for $\xi$ has been presented in Appendix \ref{App:BE}
along with the definition of thermal average of $E\times \sigma {\rm v}$
for a process like $\nu_R\overline{\nu_R}\rightarrow j\bar{j}$.
In Eq.\,(\ref{esigmaVeff}) the prime over $\langle{E \sigma {\rm v}}
\rangle_{\nu_R\overline{\nu_R}\rightarrow j\bar{j}}$ denotes
thermal average of $E\times \sigma {\rm v}$ normalised by
the product of equilibrium number densities of the final state particles
$j$ and $\bar{j}$ respectively. The Feynman diagrams
for the thermalisation of $\nu_R$ within the dark sector with $\psi$
and $\phi$ are shown in Fig.\,\ref{Fig:feyn_dia1}. Moreover, the same
interactions (interchanging the initial and the final states)
are involved in the freeze-out process of dark matter $\psi$. As a
result we obtain strong constraint on dark matter phenomenology from
the CMB bound on $\Delta{N}_{\rm eff}$ and
it has been shown in Section \ref{sec:res}.
Before going to the next section on $\Delta{N}_{\rm eff}$,
we now discuss on the DM-nucleus scattering cross section
in the remaining part of this section.  

\begin{figure}[h!]
\centering
\includegraphics[height=5.0cm,width=12cm]
{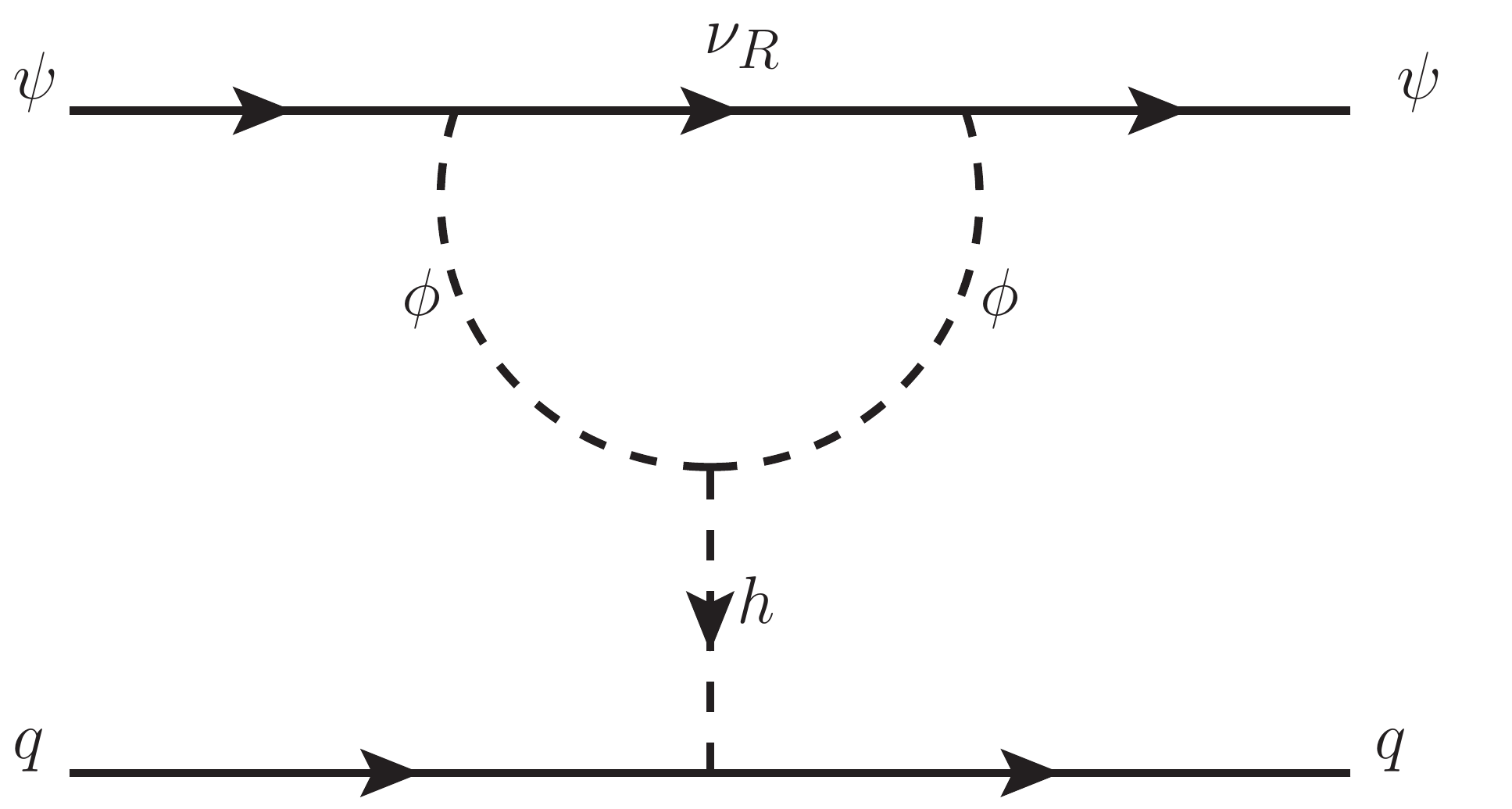}
\caption{Feynman diagram for the scattering of
$\psi$ at direct detection
experiments.}
\label{Fig:feyn_dia_dd}
\end{figure}
Although our dark matter candidate does not have any direct coupling
with the SM fields, it can still scatter off the detector nucleus
efficiently depending on the portal coupling $\lambda_{H\phi}$,
the new Yukawa coupling $y_{\phi}$ and the mass $M_{\psi}$.
The scalar mediated spin independent DM-nucleon scattering occurs at one loop
level, where the complex scalar $\phi$ and $\nu_R$ are running into
the loop and the SM Higgs is playing the role of the mediator. The Feynman
diagram of this scattering is shown in Fig.\,\ref{Fig:feyn_dia_dd}. The
spin independent scattering cross section between $\psi$ and nucleon ($N$)
is given by,
\begin{eqnarray}
\sigma_{\rm SI} = \left(\dfrac{m_N}{v}\right)^2\,\dfrac{\mu_{\psi N}^2\, 
g_{\overline{\psi}\psi\,h}^2}{\pi\,m_{h}^4}\,f_{N}^2\,,
\label{sigmaSI}
\end{eqnarray}
where, $\mu_{\psi\, N}=\frac{M_{\psi} m_{N}}{M_{\psi}+m_{N}}$ is
the reduced mass of DM-nucleon system with $m_N$ being
the mass of nucleon. The Higgs-nucleon
coupling is proportional to the factor $f_{N} \simeq 0.3$ which
depends on the quark content within a nucleon for each
quark flavour \cite{Cline:2013gha}. The effective vertex
factor for the trilinear interaction vertex $\overline{\psi}\psi h$
is denoted by $g_{\overline{\psi}\psi\,h}$ and its expression
in the low momentum transfer limit ($t << M^2_{\psi}$) is given by,
\begin{eqnarray}
g_{\overline{\psi}\,\psi\, h}= \frac{i}{16\pi^2}\times 
\frac{y_\phi^2 \lambda_{H\phi}\, v}{M_{\psi}}\times 
\left[1+ \left(\frac{M_{\phi}^2}{M_{\psi}^2} -1\right) 
{\rm ln} \left(1-\frac{M_{\psi}^2}{M_{\phi}^2} \right) \right] \,. 
\label{gpsipsih}
\end{eqnarray} 
We have computed the spin-independent scattering cross section
using Eqs.\,((\ref{sigmaSI}) and (\ref{gpsipsih})) and have found the allowed
parameter space after comparing with the latest data
from XENON1T experiment \cite{Aprile:2018dbl} in Section \ref{sec:res}. 
\section{Contribution to $N_{\rm eff}$ due to the Dirac nature of neutrinos}
\label{sec:delNeff}
As we discussed earlier, we have considered one of the
minimal extensions of the SM here to address the dark matter and
light sub-eV scale Dirac neutrinos. As a result, the right chiral parts ($\nu_R$)
are as light as the corresponding left chiral counterparts ($\nu_L$) and
thus we are bound to get additional contributions to $N_{\rm eff}$.
Consequently, the present bound on $\Delta{N_{\rm eff}}$
constrain the interaction strength of $\nu^{\alpha}_R$
and hence the decoupling from plasma prior to the onset of BBN.
Most importantly, in the present case, $\nu^{\alpha}_R$ thermalises
with the SM bath through its interaction with the
complex scalar $\phi$ and the dark matter candidate $\psi$ 
as the sub-eV scale neutrino masses require the corresponding
Yukawa couplings with the SM leptons and SM Higgs to be
as minuscule as $\sim 10^{-12}$. The decoupling of
$\nu^\alpha_{\rm R}$ is triggered as soon the kinetic
equilibrium between $\phi$ and the SM bath is lost. Thereafter,
all three $\mathbb{Z}_4$ odd species maintain a local
thermal equilibrium among themselves by virtue of the
new Yukawa interaction. As we have seen from
Fig.\,\,\ref{Fig:scattering_rate}, the kinetic decoupling of
$\phi$ essentially depends on the portal coupling $\lambda_{H\phi}$
and the mass $M_{\phi}$. This results in a lower bound
on the mass of dark matter indirectly as we need $M_{\psi} < M_{\phi} + m_{\nu}$
for a stable dark matter candidate $\psi$.
Note that DM coupling to the SM neutrinos have been studied
in the context of BBN and CMB constraints in earlier works,
see \cite{Nollett:2014lwa} for example. Typically sub-GeV thermal
DM gets constrained from such bounds if they have sizeable couplings
to SM neutrinos. As we will see in our work, DM with much heavier mass
range ($M_{\psi}\geq 1$ GeV) can also be constrained if they have couplings with light Dirac neutrinos. 
 
The additional contributions to $N_{\rm eff}$ at the time of
CMB formation due to three right handed neutrinos can be obtained
using the definition of $N_{\rm eff}$ (Eq.\,(\ref{def_neff})) as
\begin{eqnarray}
\Delta{N_{\rm eff}} &=& 
\dfrac{\sum_{\alpha}\varrho_{\nu^{\alpha}_R}}{\varrho_{\nu_L}}\,,\nonumber \\
&=&3 \times \dfrac{\varrho_{\nu_R}}{\varrho_{\nu_L}}\,, \nonumber \\
&=& 3 \times \left(\dfrac{T_{\nu_R}}{T_{\nu_L}} \right)^4\bigg|_{T_{\rm CMB}}\,,
\label{del_neff}
\end{eqnarray}
where $T_{\rm CMB}\simeq 0.26$ eV
is the photon temperature at the time of the CMB formation.
In the above, a couple of assumptions have been used.
Firstly, we have made a simplifying assumption
that all three right handed neutrinos have
same couplings for the new Yukawa interaction and this makes
the behaviour of all $\nu_R$s identical. Accordingly, we have
substituted $\sum_{\alpha}\varrho_{\nu^{\alpha}_R} =
3\times \varrho_{\nu_R}$ with $\varrho_{\nu_R}$ being the
energy density of a single species of right handed neutrino.
Moreover, we have used the equilibrium distribution
function for the right handed neutrinos throughout its
cosmological evolution. This is indeed the situation
for a massless species (e.g. photon), which was once in equilibrium
preserves its distribution function even after decoupling \cite{Dolgov:2002wy}.
The similar situation would be valid for a massive species
like neutrinos if they decouple from the plasma containing
$e^{\pm}$ and $\gamma$ instantly. In reality, this is not
the case and hence spectral distortion in the
distribution function is inevitable after decoupling. However,
the amplitude of distortion is not much significant and
for the left handed $\nu_e$ it has been shown in \cite{Dolgov:2002wy}
that $\left|\dfrac{\delta f}{f^{eq}_{\nu_e}}\right| \sim 10^{-4}$
for $E \sim T$ and $\delta f = f_{\nu_e} - f^{eq}_{\nu_e}$.
Therefore, we have neglected this small distortion in the distribution
function and consider $\varrho_{\nu_R} \propto T^4_{\nu_R}$ even
after decoupling\footnote{We have calculated the distribution
function of $\nu_R$ taking into account the decay of
$\phi \rightarrow \psi + \overline{\nu_R}$ after the decoupling
of $\nu_R$ from the SM bath. We have found that
the spectral distortion in distribution function for $T<T_{dec}$
is inadequate to produce any observable deviation in $\varrho_{\nu_R}$
and $\mathsf{n}_{\nu_R}$ from their respective equilibrium values.}.
Furthermore, for $T_{\nu_R} << T_{dec}$ when
the annihilation rate $\Gamma_{\nu^\alpha_{R}
\overline{\nu^\alpha_{R}} \rightarrow j \bar{j}}
<< \mathcal{H}$, the energy density of $\nu_R$ redshifts
as $1/{a(t)}^4$ with $a(t)$ being the cosmic scale factor
in the Friedmann-Lemaitre-Robertson-Walker (FLRW) metric. As a result, the ratio $T_{\nu_R}/T_{\nu_L}$
remains unaffected after the decoupling of $\nu_L$ as
the temperature of $\nu_L$, after $T\sim 1$ MeV,
behaves identically with the scale factor as that of $T_{\nu_R}$.
Consequently, we do not need to compute the ratio of
$T_{\nu_R}$ and $T_{\nu_L}$ at $T=T_{\rm CMB}$. Instead,
the ratio evaluated at a much larger temperature
i.e. just before the decoupling of $\nu_L$ ($T>T^{\rm dec}_{\nu_L}>>T_{\rm CMB}$)
is sufficient to determine $\Delta{N_{\rm eff}}$ at $T_{\rm CMB}$. 
Therefore, one can rewrite the expression
of $\Delta{N_{\rm eff}}$ in Eq.\,(\ref{del_neff}) as
\begin{eqnarray}
\Delta{N_{\rm eff}} &=& 3\times
\left(\dfrac{T_{{\nu}_R}}{T_{\nu_L}}\right)^4
\bigg|_{T>T^{\rm dec}_{\nu_L}}\,,\nonumber \\
&=& 3 \times \left(\dfrac{T_{{\nu}_R}}{T}\right)^4
\bigg|_{T>T^{\rm dec}_{\nu_L}}\,, \nonumber \\
&=& 3 \times \xi^4\bigg|_{T>T^{\rm dec}_{\nu_L}}\,,
\label{delNeff_final}
\end{eqnarray}
where $\xi=\dfrac{T_{\nu_R}}{T}$ as defined earlier in
Section \ref{sec:DM&RD}. In the last but one step, we have
replaced $T_{\nu_L}$ by the photon temperature $T$
as before decoupling both $\nu_L$ and photon
share a common temperature. 
\section{Numerical Results}
\label{sec:res}
\begin{figure}[h!]
\includegraphics[height=6cm,width=8.0cm,angle=0]{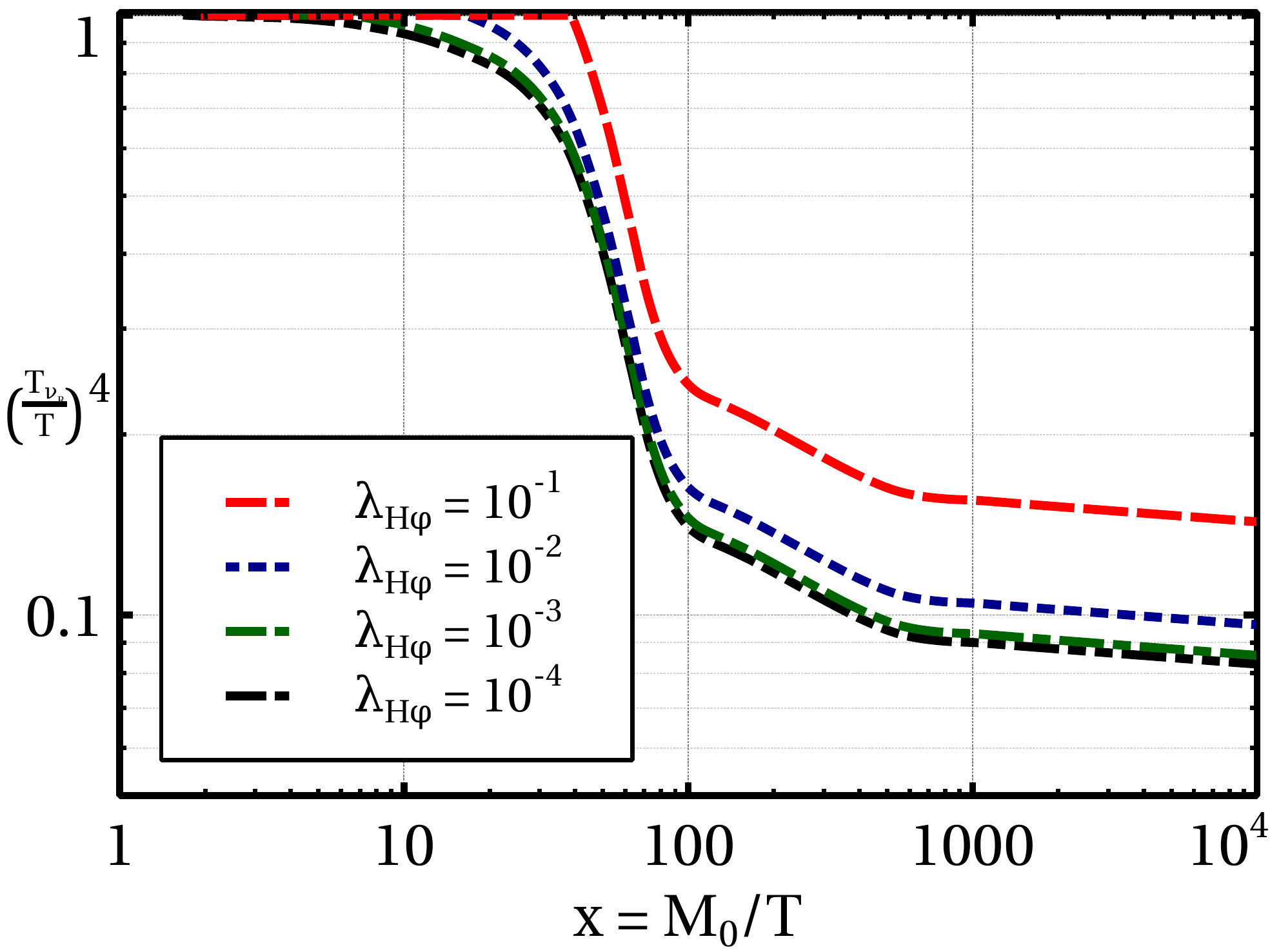}
\includegraphics[height=6cm,width=8.0cm,angle=0]{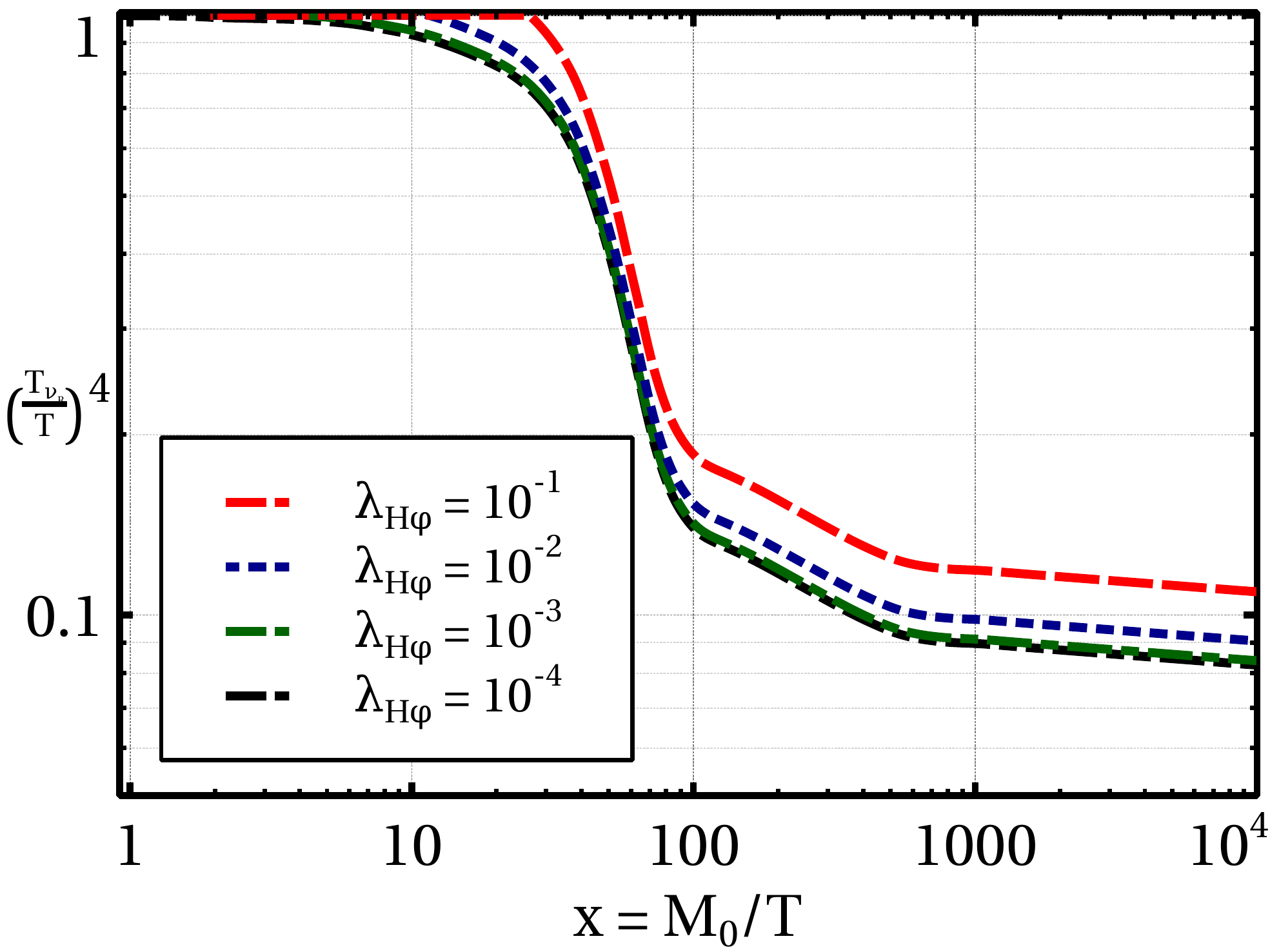}
\includegraphics[height=6cm,width=8.0cm,angle=0]{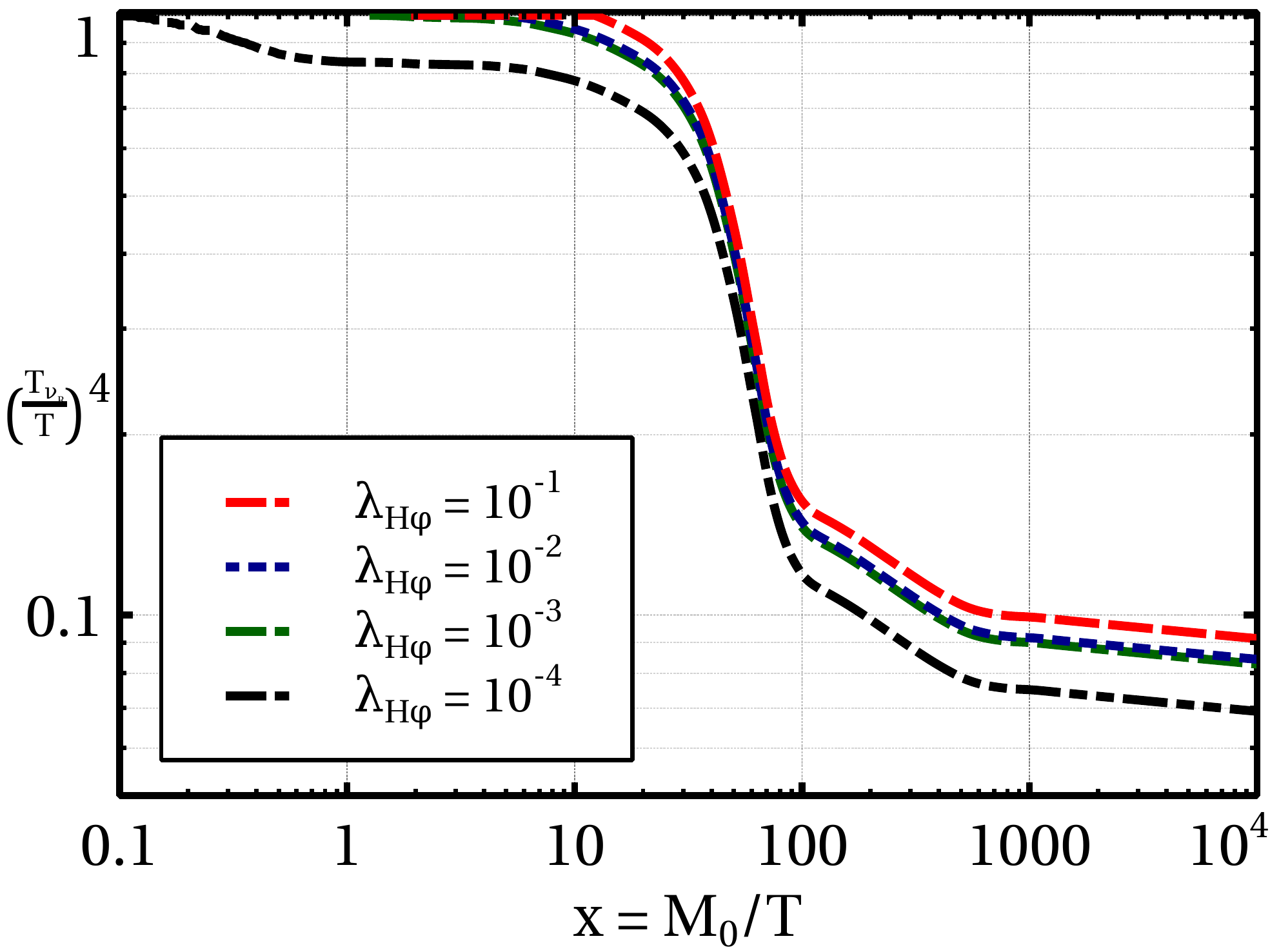}
\caption{Variation of $\left(\frac{T_{\nu_R}}{T}\right)^4$ with
$x$ for three different benchmark points
such as $M_{\phi}=25$ GeV, $M_{\psi}=5$ GeV (the left plot in upper panel),
$M_{\phi}=50$ GeV, $M_{\psi}=5$ GeV (the right plot in upper panel) and
$M_{\phi}=250$ GeV, $M_{\psi}=5$ GeV (the plot in lower panel). In the
upper panel $T$ varies between 10 GeV to 1 MeV while in the
lower panel we have 100 GeV $\leq T\leq$ 1 MeV. Moreover, in all
three plots we have kept the new Yukawa coupling $y_{\phi}$
fixed at $0.25$.}
\label{Fig:xi-vs-x}
\end{figure}

\begin{figure}[h!]
\includegraphics[height=8cm,width=6.0cm,angle=-90]{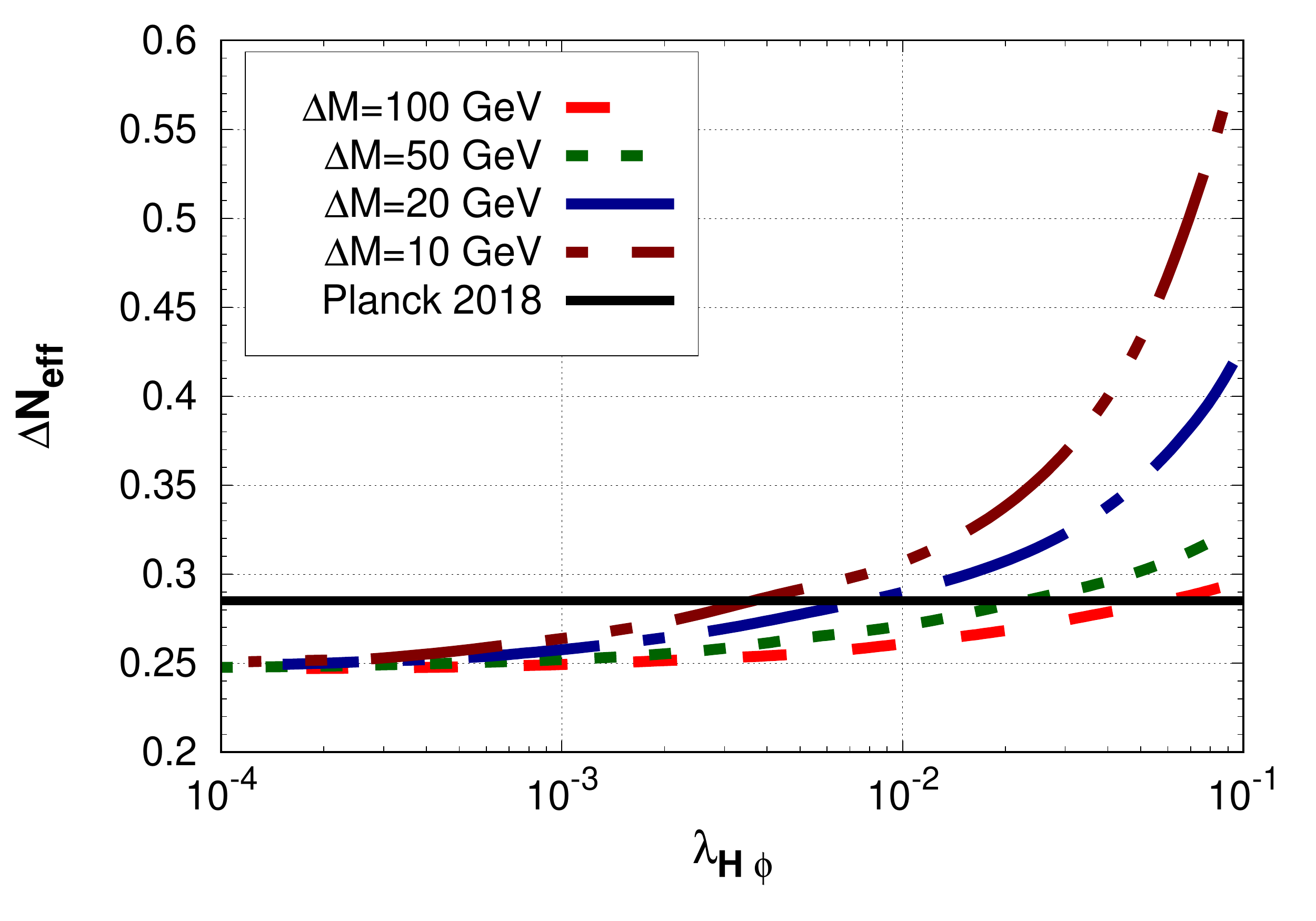}
\includegraphics[height=8cm,width=6.0cm,angle=-90]{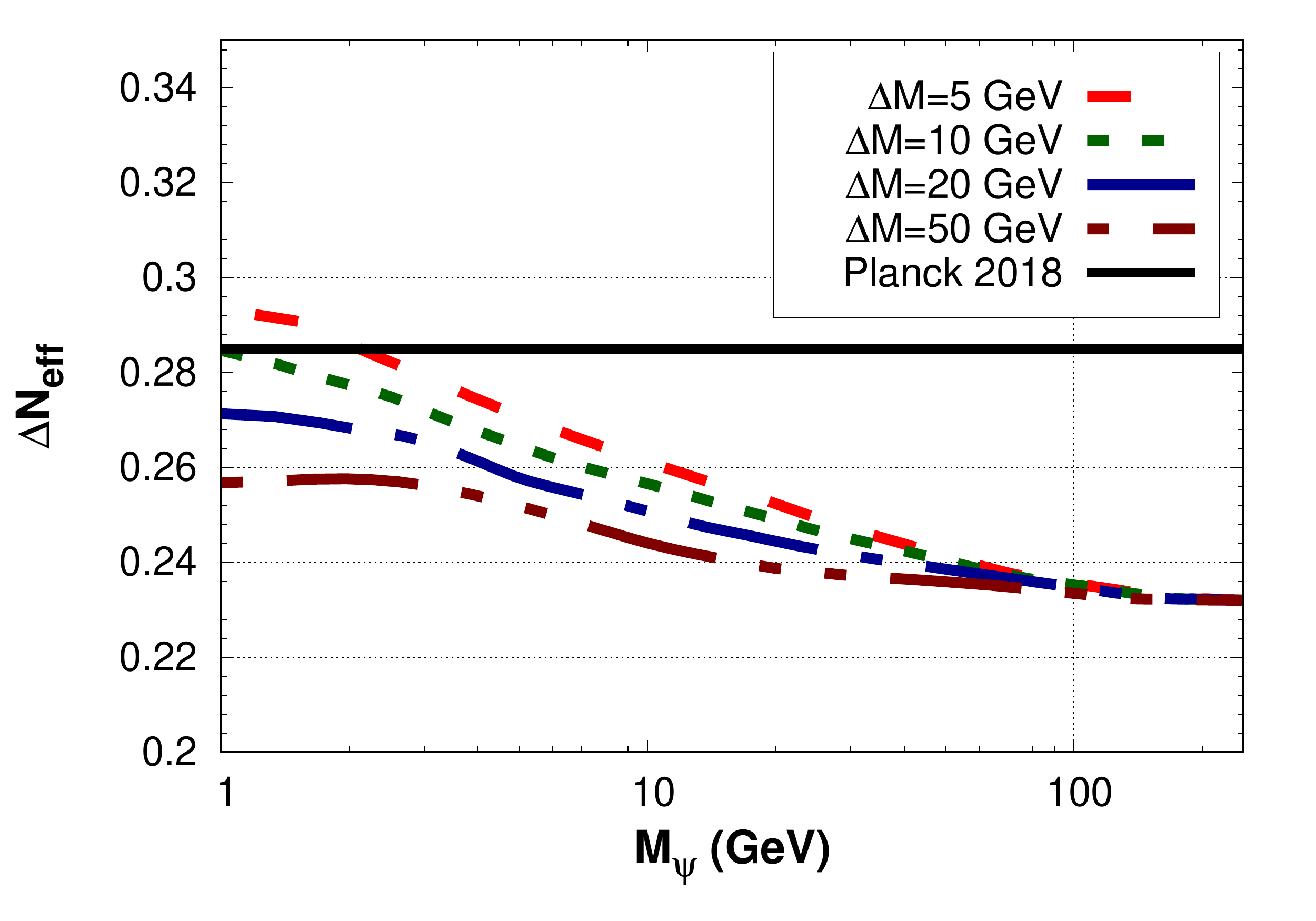}
\includegraphics[height=10cm,width=7.0cm,angle=-90]{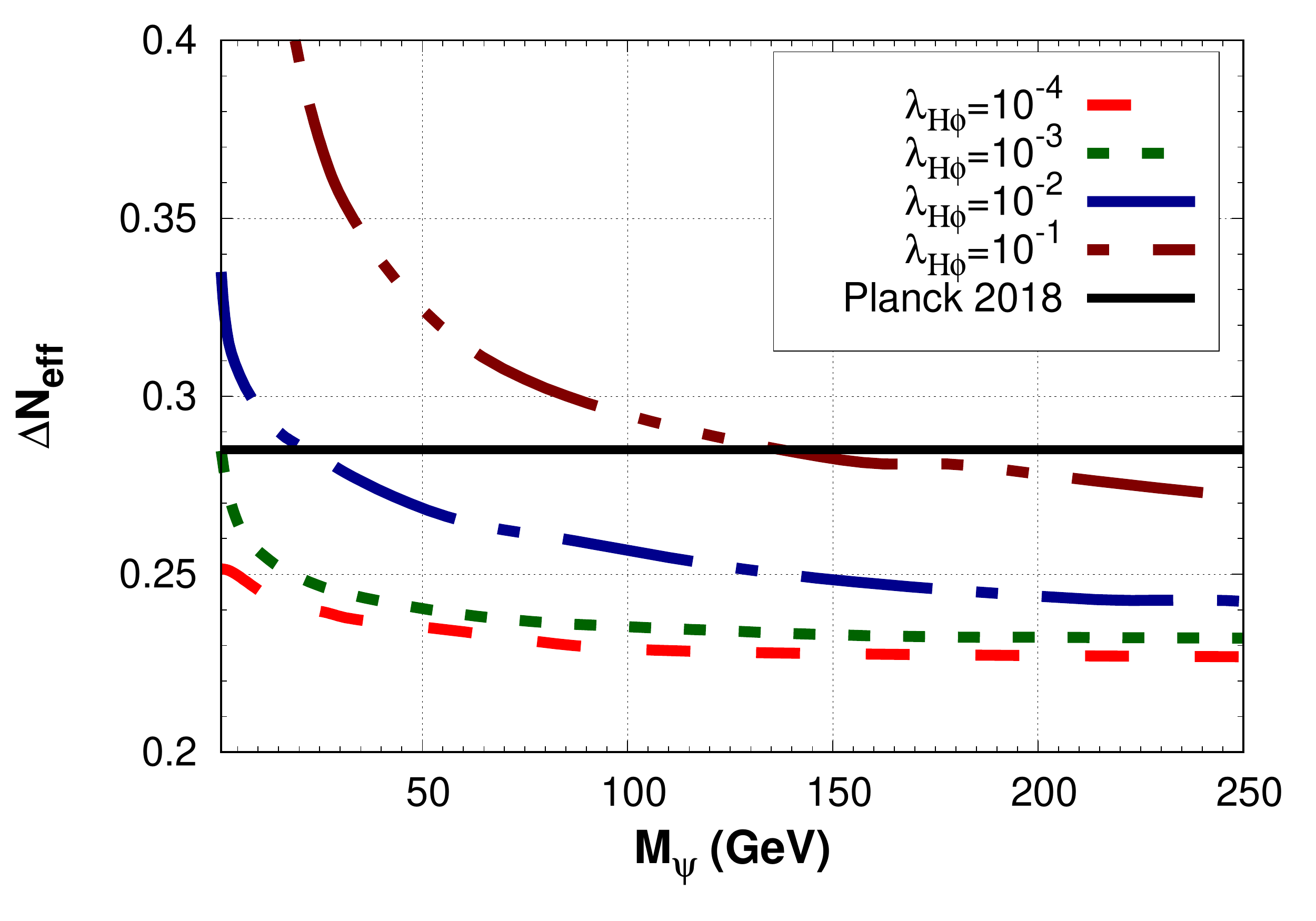}
\caption{Variation of $\Delta{N}_{\rm eff}$ with relevant model parameters.
The black solid line in each plot indicates the $2\sigma$ upper limit
of $\Delta{N}_{\rm eff}$ ($\Delta{N}_{\rm eff}=0.285$) from
the Planck 2018 data. In addition, all three plots are drawn
for $y_{\phi}=0.2$ while the portal coupling $\lambda_{H\phi}=10^{-3}$ for
the plot at top-right panel and the mass splitting $\Delta{M}=10$ GeV
for the plot at the bottom panel respectively.}
\label{Fig:delNeff-vs-model_paras}
\end{figure}

\begin{figure}[h!]
\centering
\subfigure[Variation of $\Delta{N}_{\rm eff}$ with $\lambda_{H\phi}$]{
\includegraphics[height=6.0cm,width=8.0cm,angle=0]{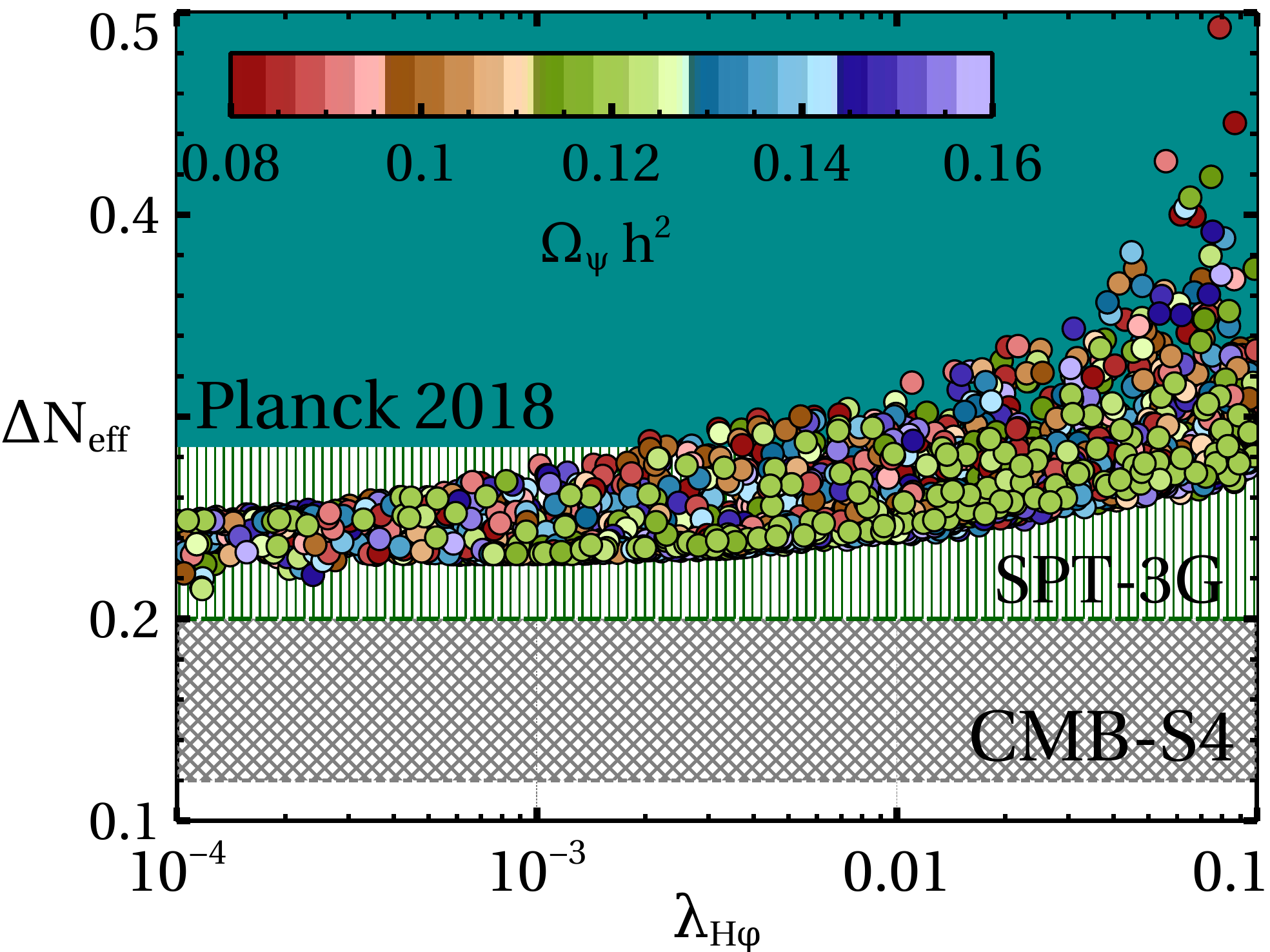}}
\subfigure[Variation of $\Delta{N}_{\rm eff}$ with $T_{dec}$]{
\includegraphics[height=6.0cm,width=8.0cm,angle=0]{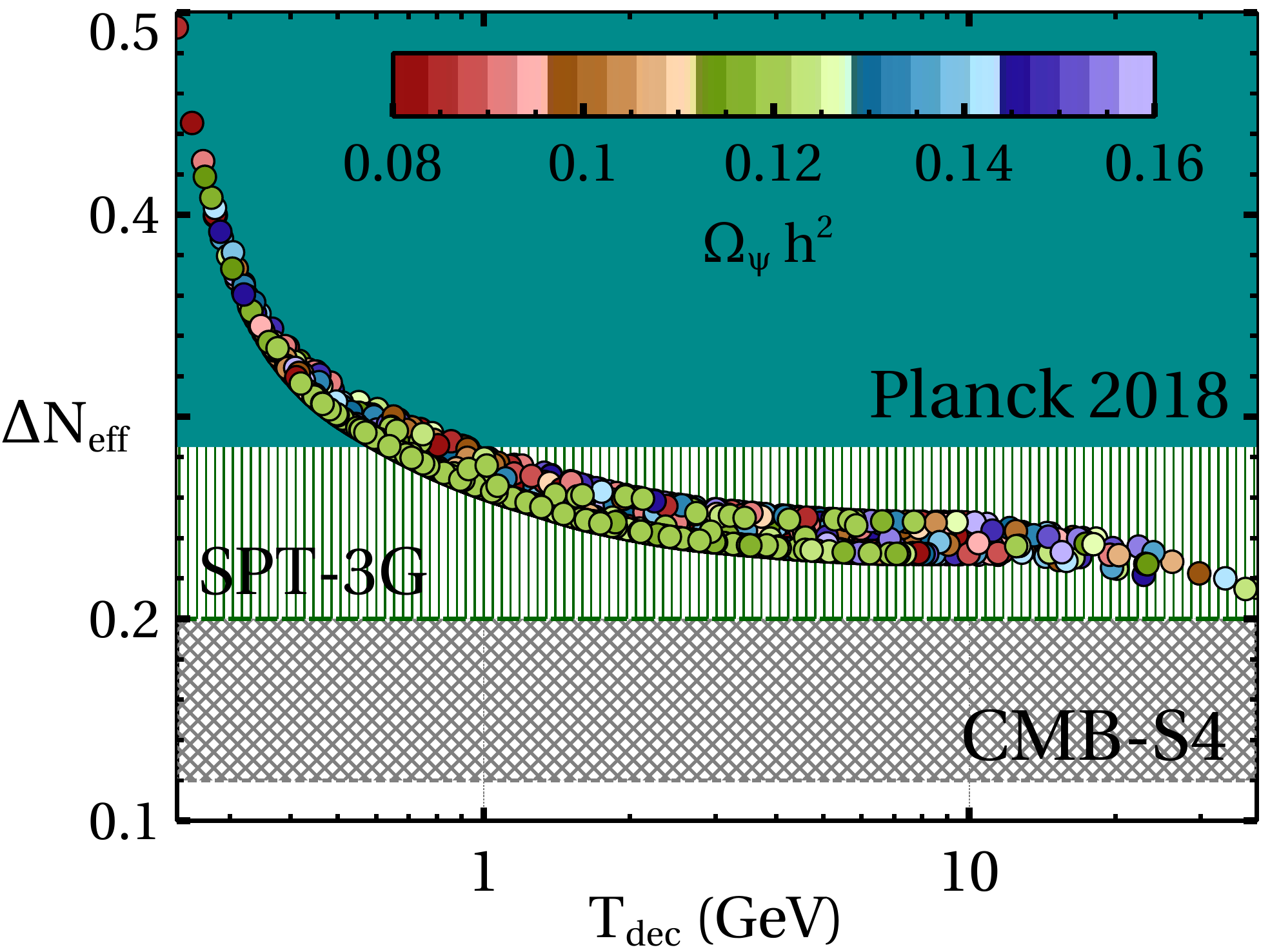}}
\subfigure[Variation of $\Delta{N}_{\rm eff}$ with $M_{\psi}$]{
\includegraphics[height=7.0cm,width=10.0cm,angle=0]{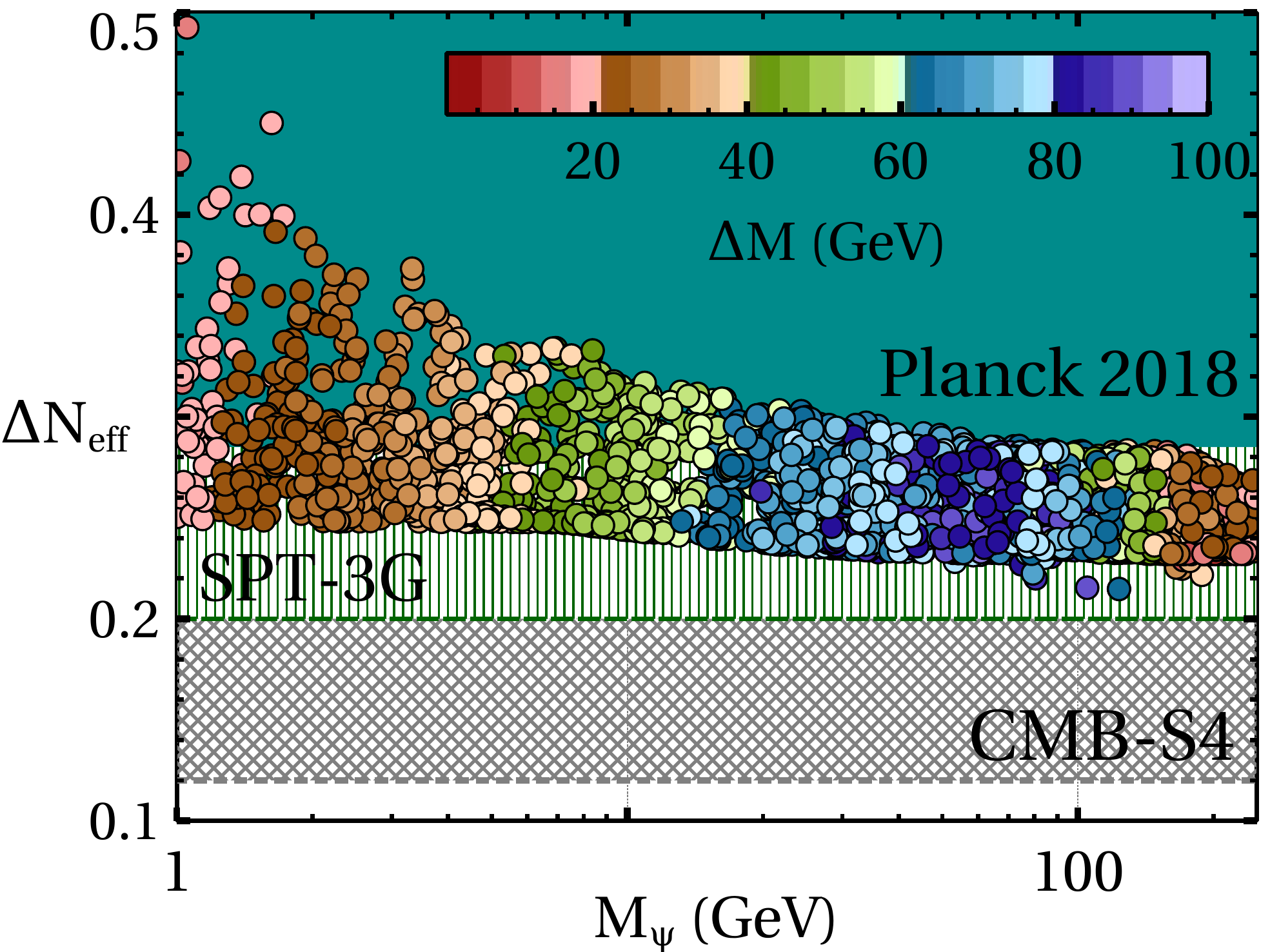}}
\caption{All three plots are generated by varying the model
parameters in the range as mentioned in Eq.\,\,(\ref{para-ranges}). In
each plot the dark cyan region is the current $2\sigma$ upper limit
on $\Delta{N_{\rm eff}}$ from the Planck 2018 data while sensitivities
of the two upcoming CMB experiments are also presented for comparison.}
\label{Fig:scan}
\end{figure}

\begin{figure}[h!]
\includegraphics[height=8.0cm,width=12.0cm,angle=0]{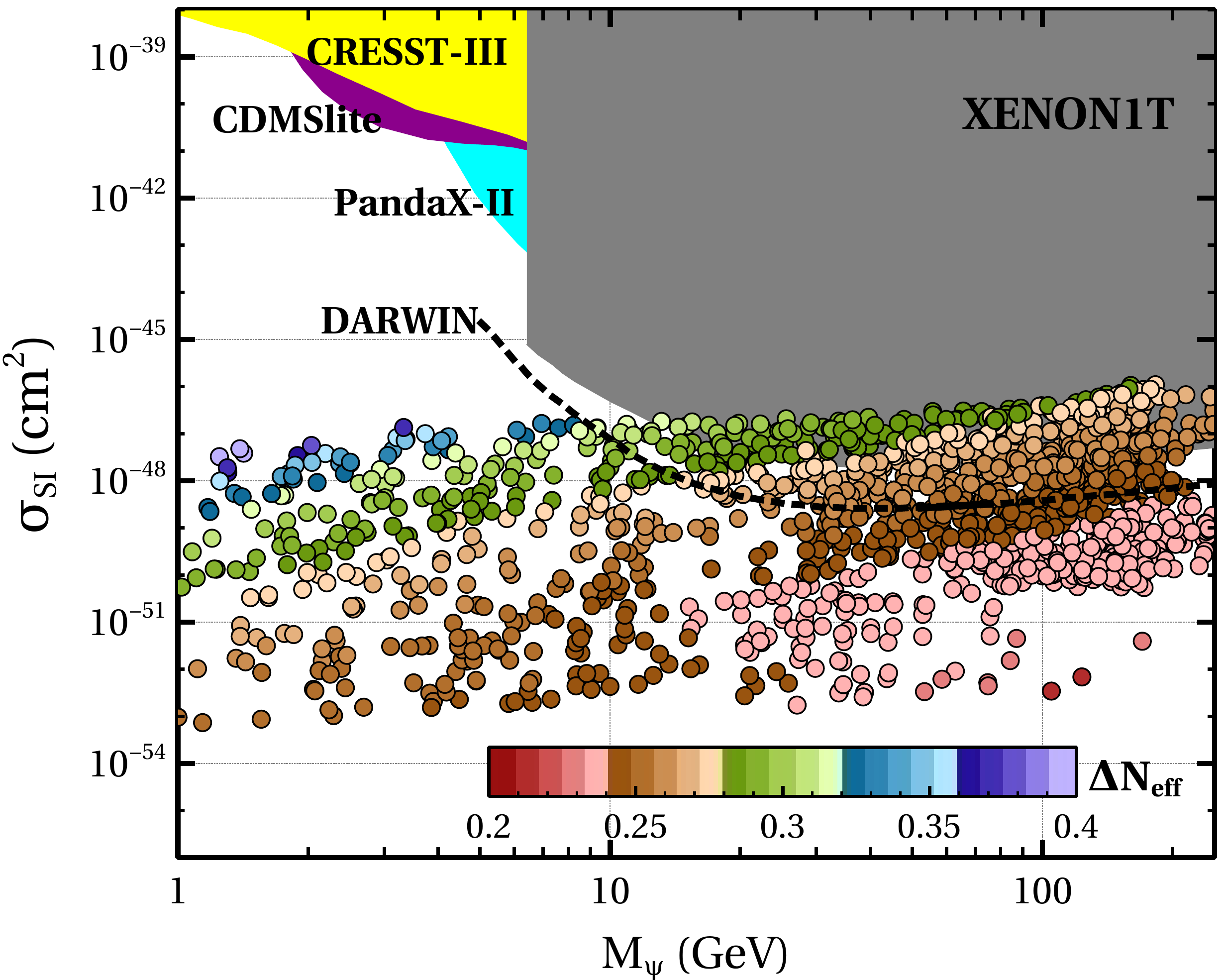}
\caption{Comparison of spin independent scattering cross section of
$\psi$ obtain from the present model with existing as well as projected
bounds from various direct detection experiments. All the points in
$\sigma_{\rm SI}-M_{\psi}$ plane satisfy relic density bound in $3\sigma$
range i.e. $0.117\leq \Omega_{\psi} h^2 \leq 0.123$. The colour bar indicates
corresponding values of $\Delta{N}_{\rm eff}$.}
\label{Fig:DD_bounds}
\end{figure}

In this section we will present our numerical results. Our principal
goal is to find the relic density of $\psi$ and the contribution
of $\nu_R$s to $N_{\rm eff}$. As we have mentioned earlier that before the kinetic
decoupling of $\phi$ ($T\geq T_{dec}$) from the SM bath,
both the sectors have a common temperature and hence $\xi=1$.
In this regime, to obtain the dark matter abundance we
require to solve the Eq.\,\,(\ref{BEy_T_gt_Tdec}) only and
it has been done using the package \texttt{micrOMEGAs} \cite{Belanger:2014vza},
where the model information has been provided to \texttt{micrOMEGAs}
using the package \texttt{FeynRules} \cite{Alloul:2013bka}.
The output of \texttt{micrOMEGAs} at $T=T_{\rm dec}$ has
been used as an input for the second regime ($T<T_{\rm dec}$).
In this regime, besides the Boltzmann equation for $Y$ given
in Eq.\,\,(\ref{BEy_T_lt_Tdec}), we need to solve another
Boltzmann equation for $\xi$ (or equivalently for $T_{\nu_R}$)
also as here $T_{\nu_R} \neq T$. The Boltzmann equation describing
the variation of $\xi$ with $x$ (inverse of $T$)
is given in Eq.\,\,(\ref{BExi}). We have solved the two
coupled differential equations numerically using our
own codes. The expressions for relevant annihilation cross sections involving
dark sector particles, which are required for solving
Eq.\,\,(\ref{BEy_T_lt_Tdec}) and Eq.\,\,(\ref{BExi}) numerically, 
are given in Appendix \ref{App:2to2xsections}. The results
are presented in Figs.\,\,\ref{Fig:xi-vs-x}-\ref{Fig:DD_bounds}. 

In Fig.\,\,\ref{Fig:xi-vs-x} we have shown the variation
of $\xi^4$ with $x$ for three different sets of dark sector
particles masses. In this work we have fixed $M_0$ at 10 GeV.
In the left plot at upper panel, we have depicted
how $\xi^4$ changes with $x$ for $M_{\psi}=5$ GeV and
$M_{\phi}=25$ GeV. In this plot we have considered four
different values of portal couplings such as $\lambda_{H\phi}=10^{-1}$
(red dashed line), $10^{-2}$ (blue dotted line),
$10^{-3}$ (green dash-dot line) and $10^{-4}$ (black dash-dot-dot line)
respectively. One can clearly observe that the era at which deviation
of $\xi$ from unity occurs gets delayed as we increase the portal coupling
$\lambda_{H{\phi}}$. This is primarily due to the reason that the
higher values of $\lambda_{H\phi}$ prolonged the kinetic equilibrium
between the dark sector and the SM bath. For example, the point
of deviation of $\xi$ from unity shifts from $x=40$ ($T=0.25$ GeV)
to $x=2$ ($T=5$ GeV) when the portal coupling is decreased by
three orders of magnitude from $\lambda_{H\phi} =10^{-1}$.
Once $\xi$ departs from unity, it continues to decrease
with $T$ and there is a sharp decrease of $\xi$
between $x\sim 10$ ($T\sim 1$ GeV) and
$x\sim 100$ ($T\sim 100$ MeV). This happens mainly due to the
term $\beta$ in the left hand side of Eq.\,(\ref{BExi}),
which is approximately equal to unity at very
high and low temperatures and gets a peak at
$T\sim 150$ MeV (around the QCD phase transition temperature)
as all effective DOFs ($g_{\rho}$, $g_s$ and $g^{1/2}_{\star}$)
diminish sharply during this period \cite{Husdal:2016haj}.
Thereafter, as $T$ reduces further and reaches up to a few MeV range,
the quantity $\beta$ again revives to almost unity and $\xi$ becomes
independent of $T$. Now, comparing plots for different $\lambda_{H\phi}$s
we can conclude that a delayed kinetic decoupling leads to
higher values of $\xi$ at $T=1$ MeV and hence a larger contribution
to $N_{\rm eff}$ (automatically follows from Eq.\,\,(\ref{delNeff_final}).
The similar nature of $\xi$ has also been observed in the $\xi$ vs $x$
plots for other two benchmark points namely $M_{\phi}=50$ GeV,
$M_{\psi}=5$ GeV (the right plot in upper panel) and $M_{\phi}=250$ GeV,
$M_{\psi}=5$ GeV (the plot in lower panel) respectively. The only change
we have noticed after comparing all three plots is that the
scenario with smallest $M_{\phi}$ contributes maximally to
$N_{\rm eff}$ as in this case $\phi$ maintains kinetic equilibrium
up to a lowest possible temperature.   

In Fig.\,\,\ref{Fig:delNeff-vs-model_paras}, we demonstrate
how $\Delta{N}_{\rm eff}$ varies with change in different model
parameters. In all three plots we have considered $y_{\phi}=0.2$. 
In the top-left plot of Fig.\,\,\ref{Fig:delNeff-vs-model_paras},
we have shown the dependence of $\Delta{N}_{\rm eff}$ on
the portal coupling $\lambda_{H\phi}$. This plot has been
drawn for four different values of mass splitting $\Delta{M}=10$ GeV,
20 GeV, 50 GeV and 100 GeV respectively between $\phi$ and $\psi$.
Similar to the previous plots of $\xi$ in Fig.\,\,\ref{Fig:xi-vs-x},
here also we observe the same nature of $\Delta{N}_{\rm eff}$
with respect to $\lambda_{H\phi}$ and $\Delta{M}$. We find
that $\Delta{N}_{\rm eff}$ behaves oppositely with respect
these two parameters. The contribution to $N_{\rm eff}$ due
to three right handed neutrinos enhances sharply as we increase
$\lambda_{H\phi}$ further however, the large mass splitting
between dark matter and $\phi$ reduces the effect
of $\nu_R$s on $N_{\rm eff}$.      

The effect of dark sector particle masses on $\Delta{N}_{\rm eff}$
will be more understandable from the second plot in the top-right panel
of Fig.\,\,\ref{Fig:delNeff-vs-model_paras} where we have illustrated
the variation of $\Delta{N}_{\rm eff}$ with $M_{\phi}$ for
four different values of mass splitting $\Delta{M}=5$ GeV, 10 GeV, 20 GeV
and 50 GeV respectively. This plot is generated for $\lambda_{H\phi}= 10^{-3}$. 
It is clearly seen from this plot that $\Delta{N}_{\rm eff}$ rises
as we lower $M_{\psi}$ from 100 GeV to 1 GeV. Moreover, comparing
the red dashed line for $\Delta{M}=5$ GeV and the brown dash-dot-dot
line for $\Delta{M}=50$ GeV, one can easily conclude that lower
mass splitting produces larger $\Delta{N}_{\rm eff}$. 
In some cases, e.g. for $\Delta{M} = 5$ GeV and $M_{\psi}\lesssim$ 2 GeV,
the contribution of three $\nu_R$s even overshoots the $2\sigma$ upper
bound from the Planck 2018 data while other scenarios are still viable.
Nevertheless, $\Delta{N}_{\rm eff}$ becomes insensitive to both these parameters beyond $M_{\psi}\sim 100$ GeV
where all four curves are almost parallel to the X axis. This is
consistent with the approximate calculation of $\Delta{N}_{\rm eff}$
using entropy conservation where the contribution of
a relativistic species to $N_{\rm eff}$ saturates if
it decouples well above $T\sim 100$ GeV \cite{Abazajian:2019oqj}.  

Finally, in the third plot at the bottom panel, 
we have depicted the combined effects of $M_{\psi}$ and
$\lambda_{H\phi}$ on $\Delta{N}_{\rm eff}$. This plot has been generated for
$\Delta{M}=10$ GeV. From this plot it is clearly seen
that the entire considered mass range of $\psi$ i.e.
$1\,\,{\rm GeV}\leq M_{\psi}\leq 250$ GeV is allowed
by the Planck 2018 data for $\lambda_{H\phi}\leq 10^{-3}$.
However, for $\lambda_{H\phi}\geq 10^{-3}$ the low mass
region of $\psi$ is already ruled out, e.g. when $\lambda_{H\phi} = 10^{-2}$
and $10^{-1}$, the mass $M_{\psi}$ up to 20 GeV and 135 GeV respectively are
excluded by the $2\sigma$ upper limit of $\Delta{N}_{\rm eff}$. 

To understand the entire picture in a well organised
manner we have shown our parameter space allowed from
both dark matter relic density constraint and the
bound on $\Delta{N}_{\rm eff}$ in Fig.\,\,\ref{Fig:scan}. In order to obtain the parameter space we have varied
our model parameters in the following range
\begin{eqnarray}
\begin{array}{cccccc}
10^{-4} &\leq & \lambda_{H\phi} & \leq & 10^{-1}\,\,,\\ 
1.0\,\,{\rm GeV} &\leq & M_{\psi}& \leq & 250.0\,\,{\rm GeV}\,\, ,\\
1.0\,\,{\rm GeV} &\leq & \Delta{M}& \leq & 100.0\,\,{\rm  GeV}\,\,,
\label{para-ranges}
\end{array}
\end{eqnarray}
while the new Yukawa coupling $y_{\phi}$ has been kept fixed 0.2.
In the top-left panel of Fig.\,\,\ref{Fig:scan}, we
have demonstrated $\Delta{N_{\rm eff}}$ vs $\lambda_{H\phi}$
parameter space where the remaining parameters are varied
according to Eq.\,\,\ref{para-ranges}. The relic density
of dark matter candidate $\psi$ has been indicated by
the colour bar where the current $3\sigma$
range of $\Omega_{\psi}h^2$ ($0.1170\leq\Omega_{\psi} h^2\leq0.1230$)
falls within the green coloured patch. Additionally,
the present $2\sigma$ upper bound on
$\Delta{N}_{\rm eff}$ from the Planck 2018 data is shown
by the dark cyan coloured region while the sensitivities
of two upcoming CMB experiments namely SPT-3G \cite{Avva:2019hzz} and
CMB-S4 \cite{Abazajian:2016yjj} have also been indicated by green vertical lines
and gray cross lines respectively. From this plot we observe that
although the relic density of dark matter satisfies the current
bound for the entire considered range of $\lambda_{H\phi}$,
the Planck limit on $\Delta{N}_{\rm eff}$ excludes some
portion of the parameter space for higher values of
the portal coupling i.e. $\lambda_{H\phi}\geq 2\times 10^{-3}$.
Nevertheless, the remaining entire parameter space will be probed
by the upcoming experiments. 

In the top-right panel of Fig.\,\,\ref{Fig:scan},
we show how $\Delta_{N_{\rm eff}}$ changes when
the decoupling temperature of $\phi$ from
the SM bath varies between 200 MeV to 30 GeV. The variation
of $T_{\rm dec}$ is due to corresponding variation of $\Delta{M}$, $M_{\psi}$
and $\lambda_{H\phi}$ in the full range as given in Eq.\,\,(\ref{para-ranges}).
This plot clearly reveals that scenarios where three right
handed neutrinos remain thermalised with the SM bath as late as
up to $T=800$ MeV are still allowed while scenarios with $T_{\rm dec}<600$ MeV
are completely ruled out for producing excess contribution to $N_{\rm eff}$. 
This is quite consistent with earlier result of Ref.\,\,\cite{Abazajian:2019oqj}.   
Moreover, in the intermediate range where $T_{\rm dec}$ lies between 800 MeV to 600 MeV
are partially allowed depending on the values of relevant parameters
namely $\Delta{M}$, $M_{\psi}$ and $\lambda_{H\phi}$. In the bottom
panel we have presented $\Delta{N}_{\rm eff}$ vs $M_{\psi}$ parameter
space, where the mass of other dark sector particle $\phi$ has
been indicated in terms of $\Delta{M}$ using colour bar. From this
plot one can easily notice that an enhanced effect of $\nu_R$s
to $N_{\rm eff}$ is obtained for smaller values of $M_{\psi}$
and $\Delta{M}$. Moreover, here also some portion of parameter
space is already excluded by the Planck 2018 data and the
remaining parameter space is well within the reach of
upcoming experiments.

As we already mentioned, our dark matter candidate $\psi$ has no direct coupling
with the SM fields. Nevertheless, $\psi$ can still scatter off the detector
nucleus at radiative level and thereby it may produce observable signal at the direct detection experiments.
The scattering occurs at one loop level as shown in Fig.\,\,\ref{Fig:feyn_dia_dd}.
We have computed the spin independent elastic scattering
$\psi N \rightarrow \psi N$ using the expression of $\sigma_{\rm SI}$
given in Eq.\,\,(\ref{sigmaSI}). The result is shown in Fig.\,\,\ref{Fig:DD_bounds}.
Moreover, to compare our result with the existing experimental
upper bounds on spin independent dark matter-nucleon cross sections,
we have shown upper limits from XENON1T \cite{Aprile:2018dbl},
PandaX-II \cite{Tan:2016zwf}, CDMSlite \cite{Agnese:2017jvy}
and CRESST-III \cite{Abdelhameed:2019hmk} experiments in the same plot.   
We find that in spite of being a loop suppressed scattering
process, the high dark matter mass region ($M_{\psi}>10$ GeV) 
is being probed by XENON1T experiment and it has already
excluded a small portion of the parameter space. Moreover,
most of the excluded parameter space is also disallowed by
the $2\sigma$ upper limit on $\Delta{N}_{\rm eff}$ from the Planck 2018
data. The future direct detection experiment
like DARWIN \cite{Aalbers:2016jon} (projected limit has
been shown by the black dashed line) will have the sensitivity
to probe the remaining parameter space in the high mass region
which is still allowed by the $2\sigma$ bound
on $\Delta{N}_{\rm eff}$ and lying just above the ``neutrino-floor'',
a region dominated by the coherent elastic neutrino-nucleus scattering.
However, for the lighter dark matter masses ($M_{\psi}\lesssim 10$ GeV)
the present direct detection experiments are not sensitive enough  
and thus the entire relic density satisfied parameter
space (in $3\sigma$ range), in this regime, remains
far beyond the reach of current and upcoming experiments as well. Nonetheless,
the low mass region is extremely sensitive to be probed by the
CMB experiments measuring $N_{\rm eff}$. From this plot, one
can easily notice that some of the allowed parameter space
in $\sigma_{\rm SI}-M_{\psi}$ plane where $\sigma_{\rm SI}$
is as low as $\sim 10^{-48}$ cm$^2$ to
$\sim 10^{-50}$ cm$^2$ depending on $M_{\psi}$
(blue, cyan and green coloured points) has already been ruled out by the
current $2\sigma$ upper bound on $\Delta{N}_{\rm eff}$. The remaining parameter space
for the entire considered range of $M_{\psi}$ is well
within the sensitivities of the upcoming CMB experiments
like CMB-S4 \cite{Abazajian:2016yjj},
SPT-3G \cite{Benson:2014qhw}, Simons Observatory \cite{Ade:2018sbj}
etc. Therefore, these next
generation CMB experiments will be able to either
validate or falsify this kind of neutrino portal
dark matter scenario in very near future.

We would like to note in passing that although in this minimal model
we require extremely tiny Yukawa coupling $\sim 10^{-12}$ for obtaining
sub-eV scale Dirac neutrino masses, such fine-tuning can be avoided by introducing another
scalar doublet $H_2$ with an induced VEV $v_2$ which could in principle be
much lower than GeV scale\footnote{See \cite{Davidson:2009ha, Davidson:2010sf,Nanda:2019nqy} for earlier discussions on similar possibilities.}. In this case, by suitably rearranging
the $\mathbb{Z}_4$ charges, one can easily forbid the Dirac mass
term involving the SM Higgs doublet while the Yukawa
interaction involving $H_2$ with coupling $y_{H_2}$ is still allowed.
Therefore, due to the induced VEV $v_2$, sub-eV scale Dirac mass
of neutrinos can be generated for much larger Yukawa coupling $y_{H_2}$.
As a result, besides the interaction with $\phi$, now $\nu_R$ can also
thermalise with the SM plasma via scatterings with leptons like
$\overline{\nu_R}\nu_R\rightarrow e^+e^-, \overline{\nu_L}\nu_L$.
However, the impact of that interaction will be significant which
keeps $\nu_R$s in thermal equilibrium up to a smallest temperature.
\section{Conclusion}
\label{sec:conclu}
We have proposed a scenario where dark matter interacts
with the Standard Model particles only via light Dirac
neutrinos, leading to a Dirac neutrino portal
dark matter scenario. In a minimal setup, this requires the extension
of the SM by three right handed neutrinos, one singlet Dirac
fermionic DM ($\psi$) and one singlet complex scalar ($\phi$)
to assist the coupling of DM with right handed neutrinos through
a new Yukawa type interaction. The right handed neutrinos couple
to the SM neutrinos via usual Yukawa interaction involving
the SM Higgs doublet and acquire sub-eV scale Dirac masses
by appropriate tuning of the corresponding Yukawa couplings.
An additional $\mathbb{Z}_4$ symmetry is introduced for
stabilising the dark matter candidate ($\psi$) and at the
same time forbidding the unwanted Majorana mass of each
$\nu_R$. The complex scalar $\phi$ thermalises with the
SM bath through a portal coupling $\lambda_{H\phi}$
with the SM Higgs boson while both DM and $\nu_R$s maintain
kinetic equilibrium with the SM plasma by virtue of the new
Yukawa interaction involving $\psi$, $\nu_R$ and $\phi$.
This leads to additional thermalised relativistic degrees
of freedom in the early universe. Hence, the right handed
neutrinos must de decoupled from the SM bath well before
the decoupling era of their left handed counterparts,
otherwise there would be too much contribution to
$N_{\rm eff}$ due to three ${\nu_R}$s. The decoupling
of $\nu_R$ occurs as the kinetic equilibrium
between $\phi$ and the SM bath is lost.
Subsequently $\psi$, $\phi$ and $\nu_R$s form
a dark sector where all three species maintain
an equilibrium among themselves with a common
temperature $T_{\nu_R}$. The era of decoupling
depends mostly on two parameters namely the
portal coupling $\lambda_{H\phi}$ and the mass of
$\phi$. 

In order to find out the relic
density of $\psi$ and $\Delta{N}_{\rm eff}$
we have solved numerically two coupled Boltzmann equations,
one for the comoving density $Y$ and the rest is for
$T_{\nu_R}$. After computing the relic density of $\psi$,
we have calculated the spin independent elastic scattering
cross section ($\sigma_{\rm SI}$) between $\psi$ and nucleon
occurring at one loop level and mediated by the SM
Higgs boson. We have found that a small
portion of the high mass region ($M_{\psi}>10$ GeV) of
our model is ruled out by the current exclusion limit on $\sigma_{\rm SI}$ from
the XENON1T experiments. Interestingly, most of this region
is also excluded from the present $2\sigma$ upper
bound on $\Delta{N}_{\rm eff}$. The proposed direct
detection experiment DARWIN will be able to probe
the entire parameter space above the ``neutrino-floor''. 
On the other hand, in the low mass regime as 
the sensitivity of direct detection experiments become
less, our relic density satisfied parameter space
remains very far to be probed directly by the current as well
as upcoming experiments. However, in the low mass regime ($M_{\psi}\lesssim 10$ GeV),
depending on $\Delta{M}$ and $\lambda_{H\phi}$, we
already have some portion of the parameter space 
where $\nu_R$ remains in the thermal bath
as late as $T\leq 600$ MeV and thereby producing excess
contribution to $\Delta{N}_{\rm eff}$. This part of the parameter space
is thus excluded by the current $2\sigma$ upper limit
$\Delta{N_{\rm eff}}\leq0.285$ from the Planck 2018 data.    
The next generation CMB experiments like CMB-S4,
SPT-3G, Simons Observatory etc. will be sensitive enough
to validate/falsify the present model by
probing the entire dark matter mass region.
Therefore, although the present
direct detection experiments are not efficient enough
in the low mass regime, such low mass dark matter
scenarios can still be probed by measuring
$\Delta{N}_{\rm eff}$ and for certain
ranges of model parameters some part of
the parameter space in low mass regime
is already excluded where $\sigma_{\rm SI}$ is
as low as $\sim 10^{-48}$ cm$^2$ to $\sim 10^{-50}$ cm$^2$.
While this scenario offers a complementary way of probing such light DM scenarios via future CMB experiments, the active direct search strategies for low mass DM may also compete CMB experiments in near future. Additionally, possible UV completions of such minimal scenarios will also offer richer phenomenology and possibilities of linking DM and Dirac neutrinos to other problems in particle physics and cosmology (see \cite{He:2020zns} for example). We leave a detailed discussion of such possibilities to future studies.
\section{Acknowledgements}
One of the authors AB would like to thank Sougata Ganguly for 
very useful discussions at various stages of this work. He also
acknowledges the cluster computing facility at IACS (pheno-server).
\,\,DB acknowledges the support from Early Career Research Award from
DST-SERB, Government of India (reference number: ECR/2017/001873). DN
would like to acknowledge Lopamudra Mukherjee for a discussion regarding
loop-induced scattering cross section.
\appendix
\section{Derivation of the Boltzmann equation expressing evolution of $T_{\nu_R}$}
\label{App:BE}
In this appendix we will present a detail derivation of the
temperature Boltzmann equation. Our starting point will be 
the total time derivative of the phase space distribution function
$f_A(p,\,t)$ of a species ${\bf A}$ is equal to the Collision term including
all possible interactions of ${\bf A}$ i.e.  
\begin{eqnarray}
\dfrac{d}{d t}f_{\boldA}(p,\,t) = \mathcal{C}[f_{\boldA}(p,\,t)]\,
\label{BE_start}
\end{eqnarray}
where $p$ is the magnitude of three momentum $\vec{p}$
of the species ${\bf A}$. Since the linear momentum of species
red-shifted as $a^{-1}$ due to the expansion of the universe, where
$a$ is the scale factor of the FLRW metric, we have $\dfrac{\dot{p}}{p}
= -\dfrac{\dot{a}}{a}=\mathcal{H}$. Using this relation, the LHS of
the Eq.\,(\ref{BE_start}) can be written as
\begin{eqnarray}
\left(\dfrac{\partial}{\partial t}  - \mathcal{H}\,p
\dfrac{\partial}{\partial p} \right) f_{\boldA}(p,\,t) = 
\mathcal{C}[f_{\boldA}(p,\,t)]\,.
\label{BE_Liouville}
\end{eqnarray}
The quantity within bracket in the above is proportional\footnote{The
actual Liouville operator is $E$ times the operator within bracket.} to the
Liouville operator for the FLRW metric. Now, consider a specific
interaction like ${\bf A} (\vec{p}_1) + B (\vec{p}_2)
\rightarrow C (\vec{p}_3)+ D (\vec{p}_4)$, where $\vec{p}_i$s
are the three momenta and the corresponding energies are
$E_i$s. We want to calculate the moment of an operator
$\mathcal{O}(p_1)$ for the species ${\bf A}$. This can be
written as 
\begin{eqnarray}
\int \dfrac{g_1\,d^3 \vec{p}_1}{(2 \pi)^3}
\left[
\left(\dfrac{\partial}{\partial t}  - \mathcal{H}\,p_1
\dfrac{\partial}{\partial p_1} \right) f_{\boldA}(p_1,\,t)
\right]\mathcal{O}(p_1)
&=&
\int
\dfrac{g_1\,d^3 \vec{p}_1}{(2 \pi)^3}\,
\mathcal{C}[f_{\boldA}(p_1,\,t)]\,\mathcal{O}(p_1)\,.
\label{BE_gen_op}
\end{eqnarray}
Here $g_1$ is the internal degrees of freedom of ${\bf A}$.
Assuming $f_A$ vanishes at the boundary, the
LHS of the above equation can be further simplified as\footnote{In
a general $n$ to $m$ scattering for appropriate symmetry
factors see Appendix A of \cite{Biswas:2020ubd}.}
\begin{eqnarray}
&&\dfrac{d }{dt} \langle \mathcal{O}(p_1)\rangle
+ 3 \mathcal{H} \left(\langle \mathcal{O}(p_1)\rangle +
\langle \dfrac{p_1}{3}\,\dfrac{\partial}
{\partial p_1}\mathcal{O}(p_1)\rangle\right)
=
\int \prod_{\alpha=1}^{4} d\Pi_{\alpha}\,
\left(2 \pi \right)^4 \delta^4\left({\bf p}_1 + {\bf p}_2
- {\bf p}_3 - {\bf p}_4 \right) \nonumber \times \\
&& ~~~~~~~~~~~~~~~~~~~~~~
\overline{\left|\mathcal{M}\right|}^2_{{\bf A}+B\rightarrow C+D}
\left[f_{C}(p_3,t) f_{D}({p}_4,t) - 
f_{\boldA}({p}_1,t) f_{B}({p}_2,t)
\right]\mathcal{O}(p_1)\,\,,
\label{BE_general_2}
\end{eqnarray}
where 
\begin{eqnarray}
\langle \mathcal{O}(p_1)\rangle = 
\int d\Pi_1\,2E_{1}\,
f_{\boldA}({p_1},\,t)\,\mathcal{O}(p_1)\,.
\label{opeator}
\end{eqnarray}
with $d\Pi_i = \dfrac{g_i\,d^3\vec{p}_i}{(2\pi)^3\,2E_i}$ a
Lorentz invariant phase space measure while ${\bf p_i}$ denotes
the four momentum corresponding to energy $E_i$ and three momentum
$\vec{p}_i$. Moreover,
$\overline{\left|\mathcal{M}\right|}^2_{{\bf A}+B\rightarrow C+D}$
is the Lorentz invariant matrix element square averaged
over both initial and final states spins. As we want to find the
evolution of temperature of a species, therefore, in the present case 
$\mathcal{O} = E_1$. Thus, using Eq.\,(\ref{opeator}) we have
$\langle\mathcal{O}\rangle = \langle E_1 \rangle = \rho_{\boldA}$ and
$\langle \dfrac{p_1}{3}\,\dfrac{\partial}
{\partial p_1}\mathcal{O}\rangle =
\langle \dfrac{p^2_1}{3\,E_1} \rangle = P_{\boldA}$,
the energy density and the pressure of $\boldA$ respectively.
In terms of $\rho_{\boldA}$ and $P_{\boldA}$ the LHS
of Eq.\,(\ref{BE_general_2}) takes the familiar form,
\begin{eqnarray}
&&\dfrac{d}{dt} \rho_{\boldA}
+ 3 \mathcal{H} \left( \rho_{\boldA} + P_{\boldA}\right)
=
\int \prod_{\alpha=1}^{4} d\Pi_{\alpha}\,
\left(2 \pi \right)^4 \delta^4\left({\bf p}_1 + {\bf p}_2
- {\bf p}_3 - {\bf p}_4 \right) \nonumber \times \\
&& ~~~~~~~~~~~~~~~~~~~~~~
\overline{\left|\mathcal{M}\right|}^2_{{\bf A}+B\rightarrow C+D}
\left[f_{C}(p_3,t) f_{D}({p}_4,t) - 
f_{\boldA}({p}_1,t) f_{B}({p}_2,t)
\right]\,E_1\,\,.
\label{BE_general_3}
\end{eqnarray}
The RHS of the above equation can be simplified in terms of
the cross section $\sigma$ of the process ${\boldA} + B \rightarrow C +D$.
Considering equilibrium distribution function as the
Maxwell-Boltzmann distribution function, one can write the
out of equilibrium distribution function of a species $\kappa$
having energy $E_{\kappa}$ and temperature $T^\prime$ as
$f_\kappa = \dfrac{n_\kappa}{n^{eq}_\kappa} \exp(-\frac{E_{\kappa}}{T^\prime})$,
where $n_{\kappa}$ is the number density of $\kappa$ and the equilibrium
number density is $n^{eq}_{\kappa}$. Therefore, the RHS of Eq.\,(\ref{BE_general_3})
can be further simplified as
\begin{eqnarray}
\hspace{-0.2in}
\dfrac{d}{dt} \rho_{\boldA}
+ 3 \mathcal{H} \left( \rho_{\boldA} + P_{\boldA}\right)
&=&
\left[\int \prod_{\beta=1}^{2}\dfrac{g_\beta\,d^3 \vec{p}_\beta}{(2\pi)^3}
\Bigg\{\dfrac{1}{4\,E_1\,E_2} \int \prod_{\alpha=3}^{4} d\Pi_{\alpha}\,
\left(2 \pi \right)^4 \delta^4\left({\bf p}_1 + {\bf p}_2
- {\bf p}_3 - {\bf p}_4 \right)\right. \nonumber \times \\
&&
\left.
\overline{\left|\mathcal{M}\right|}^2_{{\bf A}+B\rightarrow C+D}
\Bigg\} E_1\,\exp\left(-\frac{E_1+E_2}{T^\prime}\right)\right]
\left(\dfrac{n_C}{n^{eq}_C} \dfrac{n_D}{n^{eq}_D}
- \dfrac{n_{\boldA}}{n^{eq}_{\boldA}}\dfrac{n_B}{n^{eq}_B}\right),
\label{sigmadef} \\
&=& \left[\int \prod_{\beta=1}^{2}\dfrac{g_\beta\,d^3 \vec{p}_\beta}{(2\pi)^3}\,
E_1\,\exp\left(-\frac{E_1+E_2}{T^\prime}\right)
\sigma_{{\bf A}+B\rightarrow C+D}\right]
\left(\dfrac{n_C}{n^{eq}_C} \dfrac{n_D}{n^{eq}_D}
- \dfrac{n_{\boldA}}{n^{eq}_{\boldA}}\dfrac{n_B}{n^{eq}_B}\right) \nonumber \\
&=&
\langle E\,\sigma {\rm v} \rangle_{{\bf A}+B\rightarrow C+D}
\left(n^{eq}_{\boldA}n^{eq}_{B} \dfrac{n_C}{n^{eq}_C}
\dfrac{n_D}{n^{eq}_D}- n_{\boldA}{n_B}\right)\,.
\label{BE_general_4}
\end{eqnarray}  
Where, the cross section $\sigma_{{\boldA} + B \rightarrow C +D}$
is denoted by the curly bracket in Eq.\,(\ref{sigmadef}) while
the thermal average of $E \times \sigma {\rm v}$ is defined as
\begin{eqnarray}
\langle E\,\sigma {\rm v} \rangle_{{\bf A}+B\rightarrow C+D}
= \dfrac{1}{n^{eq}_{\boldA}n^{eq}_B}
\int \prod_{\beta=1}^{2}\dfrac{g_\beta\,d^3 \vec{p}_\beta}{(2\pi)^3}\,
E_1\,\exp\left(-\frac{E_1+E_2}{T^\prime}\right)
\sigma_{{\bf A}+B\rightarrow C+D}\,.
\label{EsigmaV}
\end{eqnarray}
The expression of $\langle E\,\sigma {\rm v} \rangle_{{\bf A}+B\rightarrow C+D}$
can be further simplified by changing variables from $E_1$, $E_2$ and
the angle $\theta$ between the initial state particles to $E_{\pm} = E_1 \pm E_2$
and the Mandelstam variable $s$ respectively as given in \cite{Gondolo:1990dk}.
In this case, the integration limits for these newly defined variables are
\begin{eqnarray}
&& \left|E_{-} - E_{+}\omega\right| \leq \sqrt{\dfrac{\Lambda(s,m^2_{\boldA}, m^2_B)}{s^2}}
\sqrt{E^2_{+}-s}\,, \\
&& E_{+} \geq \sqrt{s} \,,\\
&&s\geq (m_{\boldA} + m_B)^2\,,
\end{eqnarray} 
where, the function $\Lambda (a, b,c)  = (a - b -c)^2 -4 bc$
and $\omega = \dfrac{m^2_{\boldA}-m^2_B}{s}$ (assuming $m_{\boldA}\geq m_B$).
Thereafter, following the procedure given in \cite{Gondolo:1990dk},
the six dimension integration in Eq.\,\,(\ref{EsigmaV}) reduces
to an one dimensional integration on the Mandelstam variable $s$ as
\begin{eqnarray}
\langle E\,\sigma {\rm v} \rangle_{{\bf A}+B\rightarrow C+D}
= \dfrac{g_{\boldA}g_{B}}
{n^{eq}_{\boldA}n^{eq}_B}\times \dfrac{T^\prime\,\pi^2}{(2\pi)^6}
\int_{(m_{\boldA} + m_B)^2}^{\infty} \sigma_{{\bf A}+B\rightarrow C+D}\,
\Lambda(s,m^2_{\boldA}, m^2_B)\,(1+\omega)
\,{\rm K}_2\left(\frac{\sqrt{s}}{T^\prime}\right)\,ds\,. \nonumber
\label{EsigmaV_K2_m1neqm2} \\
\end{eqnarray}
The equilibrium number density of a species $j$ obeying the Maxwell-Boltzmann
distribution and having mass $m_j$, internal degrees
of freedom $g_j$ and temperature $T^\prime$ is given by
\begin{eqnarray}
n^{eq}_{j}(T^\prime) = \dfrac{g_j\,T^\prime}{2\pi^2} m^2_j\,
{\rm K}_2\left(\frac{m_j}{T^\prime}\right)\,.
\label{neq}
\end{eqnarray}
Here, ${\rm K}_2\left(\frac{m_i}{T^\prime}\right)$ is
the modified Bessel function of second kind.
Therefore, if the initial state particles have same masses
i.e. $m_{\boldA}=m_B = m$, the above expression of
$\langle E\,\sigma {\rm v} \rangle$ reduces to the
following simpler form,
\begin{eqnarray}
\langle E\,\sigma {\rm v} \rangle_{{\bf A}+B\rightarrow C+D}
= \dfrac{1}{16\,T^\prime m^4\,\left({\rm K}_2\left(\frac{m}
{T^\prime}\right)\right)^2}
\int_{4m^2}^{\infty} \sigma_{{\bf A}+B\rightarrow C+D}\,
s\,(s-4m^2)\,{\rm K}_2\left(\frac{\sqrt{s}}{T^\prime}\right)\,ds\,.
\label{EsigmaV_K2}
\end{eqnarray}
Moreover, during the derivation of Eq.\,(\ref{BE_general_4})
we have considered same temperature $T^\prime$ for all four species
$\boldA$, $B$, $C$ and $D$. However, if the initial
and the final states particles have different temperature namely $T_i$ and $T_f$
with $T_i \neq T_f$ then the Boltzmann equation of $\rho_A$ is given by
\begin{eqnarray}
\dfrac{d}{dt} \rho_{\boldA}
+ 3 \mathcal{H} \left( \rho_{\boldA} + P_{\boldA}\right)
&=&
\langle E\,\sigma {\rm v} \rangle_{{\bf A}+B\rightarrow C+D}^{T_f}\,\,
n^{eq}_{\boldA}(T_f)n^{eq}_{B}(T_f) \dfrac{n_C(T_f)}{n^{eq}_C(T_f)}
\dfrac{n_D(T_f)}{n^{eq}_D(T_f)} \nonumber \\  
&&-\langle E\,\sigma {\rm v} \rangle_{{\bf A}+B\rightarrow C+D}^{T_i}
\,\,n_{\boldA}(T_i){n_B}(T_i)\,.
\end{eqnarray}
One can easily check that the RHS of the above equation reduces
to Eq.\,(\ref{BE_general_4}) for $T_i=T_f=T^\prime$. 

Before going to the specific case of the present model, we would
like to discuss the possible effect of a decay process like
${B}(\vec{p}_1)\rightarrow {\boldA} (\vec{p}_2)+ C(\vec{p}_3)$
to the evolution of $\rho_{\boldA}$. The collision
term contributing to the evolution
of $\rho_{\boldA}$ due to this decay has the following form,
\begin{eqnarray}
\mathcal{C}_{B\rightarrow {\boldA}+ C} &=& 
\int \prod_{\alpha=1}^{3} d\Pi_{\alpha}\,
\left(2 \pi \right)^4 \delta^4\left({\bf p}_1 - {\bf p}_2
- {\bf p}_3 \right) 
\overline{\left|\mathcal{M}\right|}^2_{{B\rightarrow {\boldA}+C}}
\left[f_{B}({p}_1,t) - f_{\boldA}(p_2,t) f_{C}({p}_3,t)
\right]\,E_2\,\,,\nonumber \\
&=& \left(\dfrac{n_B}{n^{eq}_B}-\dfrac{n_{\boldA}}{n^{eq}_{\boldA}}
\dfrac{n_C}{n^{eq}_C}\right)
\int \prod_{\alpha=1}^{3} d\Pi_{\alpha}\,
\left(2 \pi \right)^4 \delta^4\left({\bf p}_1 - {\bf p}_2
- {\bf p}_3 \right) 
\overline{\left|\mathcal{M}\right|}^2_{{B\rightarrow {\boldA}+C}}
\nonumber \times \\
&& ~~~~~~~~~~~~~~~~~~~~~
E_2 \exp\left(-\frac{E_2+E_3}{T^\prime}\right)\,.
\label{C_decay}
\end{eqnarray}
Here we have assumed CP invariance as we did earlier in Eq.\,\,(\ref{BE_general_2}).
In the last step we have used the Maxwell-Boltzmann distribution
function for all the species. Now, rearranging the collision
term one can write it as
\begin{eqnarray}
\mathcal{C}_{B\rightarrow {\boldA}+ C} &=& 
 \left(\dfrac{n_B}{n^{eq}_B}-\dfrac{n_{\boldA}}{n^{eq}_{\boldA}}
\dfrac{n_C}{n^{eq}_C}\right)\int \prod_{\alpha=2}^{3} d\Pi_{\alpha}
\,E_2 \exp\left(-\frac{E_2+E_3}{T^\prime}\right) \times \nonumber \\
&& ~~~~~~~~\left\{\int d\Pi_1 \left(2 \pi \right)^4 \delta^4\left({\bf p}_1 - {\bf p}_2
- {\bf p}_3 \right) \overline{\left|\mathcal{M}\right|}^2_{B\rightarrow{\boldA}+C}
\right\}\,.
\label{C_decay1}
\end{eqnarray}   
The term within the curly brackets is Lorentz invariant and hence,
can be evaluated in any inertial frame of reference. The most
convenient is to calculate this term in the centre of momentum frame
where 
\begin{eqnarray}
\int d\Pi_1 \left(2 \pi \right)^4 \delta^4\left({\bf p}_1 - {\bf p}_2
- {\bf p}_3 \right) \overline{\left|\mathcal{M}\right|}^2_{B\rightarrow{\boldA}+C}
=\dfrac{g_{B}\,\pi}{m_{B}}\, \delta(\sqrt{s}-m_B)\, 
\overline{\left|\mathcal{M}\right|}^2_{B\rightarrow{\boldA}+C}\,.
\end{eqnarray} 
Substituting this to the Eq.\,\,(\ref{C_decay1}), we get
\begin{eqnarray*}
\mathcal{C}_{B\rightarrow {\boldA}+ C} &=&
\left(\dfrac{n_B}{n^{eq}_B}-\dfrac{n_{\boldA}}{n^{eq}_{\boldA}}
\dfrac{n_C}{n^{eq}_C}\right)
\dfrac{g_{B}\,\pi}{m_{B}}\, 
\overline{\left|\mathcal{M}\right|}^2_{B\rightarrow{\boldA}+C}
\int \prod_{\alpha=2}^{3} d\Pi_{\alpha}
\,E_2 \exp\left(-\frac{E_2+E_3}{T^\prime}\right)
\delta(\sqrt{s}-m_B)\,,
\end{eqnarray*} 
where, we have taken $\overline{\left|\mathcal{M}\right|}^2_{B\rightarrow{\boldA}+C}$
outside the integration as the matrix amplitude square for the decay
(and also for the inverse decay as well) can be expressed in terms of
masses. Now, the six dimensional integration has the same form as
we have encountered in Eq.\,\,(\ref{EsigmaV}) except the delta function
on $\sqrt{s}$. Therefore, following the similar procedure we have found
that
\begin{eqnarray}
\int \prod_{\alpha=2}^{3} d\Pi_{\alpha}
\,E_2 \exp\left(-\frac{E_2+E_3}{T^\prime}\right)
\delta(\sqrt{s}-m_B) &=& \dfrac{g_{\boldA}g_{C}T^\prime\,m_B}{2^6\pi^4}
\sqrt{\Lambda(m^2_{B}, m^2_{\boldA}, m^2_C)} \times \nonumber \\
&&\left(1+\dfrac{m^2_{\boldA}-m^2_C}{m^2_{B}}\right)\,
{\rm K}_2\left(\frac{m_{B}}{T^\prime}\right)\,.
\end{eqnarray}
Using this, a compact form of the collision term
for the decay process ${B\rightarrow {\boldA}+ C}$ is given by,
\begin{eqnarray}
\mathcal{C}_{B\rightarrow {\boldA}+ C} &=&
\left|\mathcal{M}\right|^2_{B\rightarrow{\boldA}+C}
\dfrac{T^\prime}{2^6\pi^3}
\left(\dfrac{n_B}{n^{eq}_B}-\dfrac{n_{\boldA}}{n^{eq}_{\boldA}}
\dfrac{n_C}{n^{eq}_C}\right)
\sqrt{\Lambda(m^2_{B}, m^2_{\boldA}, m^2_C)}
\left(1+\dfrac{m^2_{\boldA}-m^2_C}{m^2_{B}}\right)\,
{\rm K}_2\left(\frac{m_{B}}{T^\prime}\right)\,.\nonumber \\
&=& \langle E\,\Gamma \rangle_{B\rightarrow{\boldA}+C}\,n_{B}^{eq} 
\left(\dfrac{n_B}{n^{eq}_B}-\dfrac{n_{\boldA}}{n^{eq}_{\boldA}}
\dfrac{n_C}{n^{eq}_C}\right)\,,
\label{C_decay_final}
\end{eqnarray}
where,
\begin{eqnarray}
\langle E\,\Gamma \rangle_{B\rightarrow{\boldA}+C}&\equiv&
\dfrac{\left|\mathcal{M}\right|^2_{B\rightarrow{\boldA}+C}}{n_{B}^{eq}}
\dfrac{T^\prime}{2^6\pi^3}
\sqrt{\Lambda(m^2_{B}, m^2_{\boldA}, m^2_C)}
\left(1+\dfrac{m^2_{\boldA}-m^2_C}{m^2_{B}}\right)\,
{\rm K}_2\left(\frac{m_{B}}{T^\prime}\right)\,,\nonumber \\
&=& \dfrac{1}{32 \pi}\dfrac{\left|\mathcal{M}\right|^2_{B\rightarrow{\boldA}+C}}{g_{B}}
\dfrac{\sqrt{\Lambda(m^2_{B}, m^2_{\boldA}, m^2_C)}}{m^2_B}
\left(1+\dfrac{m^2_{\boldA}-m^2_C}{m^2_{B}}\right)\,.
\label{Egamma}
\end{eqnarray}
In the last step we have used Eq.\,\,(\ref{neq}) for the
equilibrium number density of the species $B$.
 
Now, we will derive the evolution equation for the 
dark sector temperature $T_{\nu_R}$ starting from the
Boltzmann equation for the energy density. As we have mentioned before that
all the dark sector species after the kinetic decoupling of $\phi$ from
the SM bath have a common temperature $T_{\nu_R}$. Therefore,
to find out how $T_{\nu_R}$ evolves with time we first
need to calculate the rate of change of energy density
of $\nu_R$ within the dark sector due to scatterings and decays.
There are two possible annihilation modes of $\nu_R$
namely $\nu_R\overline{\nu}_R\rightarrow
\psi\overline{\psi}$ and $\nu_R\overline{\nu}_R
\rightarrow \phi\phi^\dagger$.
The evolution of $\rho_{\nu_R}$
can be found following Eqs.\,(\ref{BE_general_4})
and Eq.\,(\ref{C_decay_final}) as
\begin{eqnarray}
\dfrac{d}{dt} \rho_{\nu_R}
+ 3 \mathcal{H} \left( \rho_{\nu_R} + P_{\nu_R}\right)
&=&
\langle E\,\sigma {\rm v} \rangle_{
\nu_R\overline{\nu_R}\rightarrow\psi\overline{\psi}}
\left(n^{eq}_{\nu_R}n^{eq}_{\overline{\nu_R}}
\dfrac{n_\psi}{n^{eq}_\psi}\dfrac{n_{\overline{\psi}}}
{n^{eq}_{\overline{\psi}}}- n_{\nu_R}n_{\overline{\nu_R}}\right)
\nonumber \\
&+& 
\langle E\,\sigma {\rm v} \rangle_{
\nu_R\overline{\nu_R}\rightarrow\phi{\phi^\dagger}}
\left(n^{eq}_{\nu_R}n^{eq}_{\overline{\nu_R}}
\dfrac{n_\phi}{n^{eq}_\phi}\dfrac{n_{\phi^\dagger}}
{n^{eq}_{\phi^\dagger}}- n_{\nu_R}n_{\overline{\nu_R}}\right)
\nonumber \\
&+& 
\langle E\,\Gamma \rangle_{\phi^\dagger
\rightarrow{\nu_R}+\overline{\psi}}\,n_{\phi^\dagger}^{eq} 
\left(\dfrac{n_{\phi^\dagger}}{n^{eq}_{\phi^\dagger}}-
\dfrac{n_{\nu_R}}{n^{eq}_{\nu_R}}
\dfrac{n_{\overline{\psi}}}{n^{eq}_{\overline{\psi}}}\right)
\label{rhonuR}
\end{eqnarray}
Similarly, one can write a Boltzmann equation
for $\overline{\nu_R}$ also where the collision terms for scatterings
will be exactly identical to those of $\nu_R$. Therefore, the Boltzmann equation for
total energy density $\varrho_{\nu_R} = \rho_{\nu_R} + \rho_{\overline{\nu_R}}$
is given by
\begin{eqnarray}
\dfrac{d}{dt} \varrho_{\nu_R}
+ 3 \mathcal{H} \left( \varrho_{\nu_R} + 
\mathfrak{P}_{\nu_R}\right)
&=&
2\,\Bigg[\langle E\,\sigma {\rm v} \rangle_{
\nu_R\overline{\nu_R}\rightarrow\psi\overline{\psi}}
\left(n^{eq}_{\nu_R}n^{eq}_{\overline{\nu_R}}
\dfrac{n_\psi}{n^{eq}_\psi}\dfrac{n_{\overline{\psi}}}
{n^{eq}_{\overline{\psi}}}- n_{\nu_R}n_{\overline{\nu_R}}\right)
\nonumber \\
&+& 
\langle E\,\sigma {\rm v} \rangle_{
\nu_R\overline{\nu_R}\rightarrow\phi{\phi^\dagger}}
\left(n^{eq}_{\nu_R}n^{eq}_{\overline{\nu_R}}
\dfrac{n_\phi}{n^{eq}_\phi}\dfrac{n_{\phi^\dagger}}
{n^{eq}_{\phi^\dagger}}- n_{\nu_R}n_{\overline{\nu_R}}\right)
\Bigg]
\nonumber \\
&+& 
\langle E\,\Gamma \rangle_{\phi^\dagger
\rightarrow{\nu_R}+\overline{\psi}}\,n_{\phi^\dagger}^{eq} 
\left(\dfrac{n_{\phi^\dagger}}{n^{eq}_{\phi^\dagger}}-
\dfrac{n_{\nu_R}}{n^{eq}_{\nu_R}}
\dfrac{n_{\overline{\psi}}}{n^{eq}_{\overline{\psi}}}\right)
\nonumber \\
&+& 
\langle E\,\Gamma \rangle_{\phi
\rightarrow \overline{\nu_R} +\psi}\,n_{\phi}^{eq} 
\left(\dfrac{n_{\phi}}{n^{eq}_{\phi}}-
\dfrac{n_{\overline{\nu_R}}}{n^{eq}_{\overline{\nu_R}}}
\dfrac{n_{\psi}}{n^{eq}_{\psi}}\right)
\,,
\end{eqnarray}
where, $\mathfrak{P}_{\nu_R} = P_{\nu_R} + P_{\overline{\nu_R}}$ is the
total pressure. The above equation can be simplified
further if we assume that there is no asymmetry between the
number densities of particles  and antiparticles of
a species i.e. $n_{\alpha} =
n_{\bar{\alpha}} = {\mathsf{n}_{\alpha}}/2$
($\alpha = \phi,\,\psi$ and $\nu_R$) where $\mathsf{n}_{\alpha}$
is the total number density for a species $\alpha$ (including
contribution from $\bar{\alpha}$). Moreover, we assume that
the spectral distortion of $\nu_R$ even after its decoupling
is negligible similar to $\nu_L$ below $T=1$ MeV
as mentioned in \cite{Dolgov:2002wy}. Therefore,
we have taken $\mathsf{n}_{\nu_R} = 
\mathsf{n}^{eq}_{\nu_R}$ and $\varrho_{\nu_R} \propto T^4$
throughout the work (upto $T\gtrsim 1$ MeV). Therefore, the collision
term has the following form
\begin{eqnarray}
\dfrac{d}{dt} \varrho_{\nu_R}
+ 4 \mathcal{H} \varrho_{\nu_R} 
&=& 
\dfrac{(\mathsf{n}^{eq}_{\nu_R})^2}{2}
\,\Bigg[\langle E\,\sigma {\rm v} \rangle_{
\nu_R\overline{\nu_R}\rightarrow\psi\overline{\psi}}
\left(\left(
\dfrac{\mathsf{n}_\psi}{\mathsf{n}^{eq}_\psi}\right)^2
- 1\right)
+ 
\langle E\,\sigma {\rm v} \rangle_{
\nu_R\overline{\nu_R}\rightarrow\phi{\phi^\dagger}}
\left(\left(
\dfrac{\mathsf{n}_\phi}{\mathsf{n}^{eq}_\phi}
\right)^2
- 1\right)
\Bigg]
\nonumber  \\
&& ~~~~~~~~~+
\,\dfrac{\mathsf{n}_{\phi}^{eq}}{2}
\left(\langle E\,\Gamma \rangle_{\phi^\dagger
\rightarrow{\nu_R}+\overline{\psi}} +
\langle E\,\Gamma \rangle_{\phi
\rightarrow \overline{\nu_R} +\psi}\right) 
\left(\dfrac{\mathsf{n}_{\phi}}{\mathsf{n}^{eq}_{\phi}}-
\dfrac{\mathsf{n}_{\psi}}{\mathsf{n}^{eq}_{\psi}}\right)\,.
\label{BE_rhonuR_d&s}
\end{eqnarray}
In the left hand side we have used $\mathfrak{P}_{\nu_R} = \dfrac{1}{3}\varrho_{\nu_R}$ for
the relativistic species $\nu_R$. Now, if we use the relation
$\dfrac{\mathsf{n}_{\alpha}}{\mathsf{n}}\simeq
\dfrac{\mathsf{n}^{eq}_{\alpha}}{\mathsf{n}^{eq}}$ (for $\alpha=\phi,\,\psi$)
\cite{Griest:1990kh, Edsjo:1997bg} with
$\mathsf{n} = \mathsf{n}_{\psi}+ \mathsf{n}_{\phi}$ in Eq.\,\,(\ref{BE_rhonuR_d&s}),
the contribution coming from the decays of $\phi$ and $\phi^\dagger$
vanishes while the effect of scatterings survives,
\begin{eqnarray}
\dfrac{d}{dt} \varrho_{\nu_R}
+ 4 \mathcal{H} \varrho_{\nu_R} 
&=& 
\dfrac{1}{2} \left(\dfrac{\mathsf{n}^{eq}_{\nu_R}}
{\mathsf{n}^{eq}}\right)^2
\left(\langle E\,\sigma {\rm v} \rangle_{
\nu_R\overline{\nu_R}\rightarrow\psi \overline{\psi}}
+
\langle E\,\sigma {\rm v} \rangle_{
\nu_R\overline{\nu_R}\rightarrow\phi{\phi^\dagger}}
\right) \left[\mathsf{n}^2 
- (\mathsf{n}^{eq})^2\right].
\label{BE_rhonuR:app}
\end{eqnarray}
Now, substituting $\varrho_{\nu_R} = \alpha\,T^4_{\nu_R}$
\footnote{Here, for simplicity, we have used the Maxwell-Boltzmann
distribution for $\nu_R$. The energy density of $\nu_R$ becomes
proportional to $T^4_{\nu_R}$ in the limit $m_{\nu}<<T_{\nu_R}$.} with
$\alpha = g_{\nu_R}\times \dfrac{7}{8}\dfrac{\,\,\pi^2}{30}$,
$g_{\nu_R} = 2$ in Eq.\,(\ref{BE_rhonuR:app}) and
replacing time $t$ by temperature $T$
using the time-temperature relationship
$\dfrac{dT}{dt} = -\dfrac{\mathcal{H}\,T}{\beta}$ we get,
\begin{eqnarray}
\hspace{-0.5cm}
T\,\dfrac{d\xi}{dT} + (1-\beta)\xi &=&
-\dfrac{1}{2}\,\dfrac{\beta}{4\,\alpha\,\xi^3\,T^4\,\mathcal{H}}
\left(\dfrac{\mathsf{n}^{eq}_{\nu_R}}
{\mathsf{n}^{eq}}\right)^2
\left(\langle E\,\sigma {\rm v} \rangle_{
\nu_R\overline{\nu_R}\rightarrow\psi \overline{\psi}}
+
\langle E\,\sigma {\rm v} \rangle_{
\nu_R\overline{\nu_R}\rightarrow\phi{\phi^\dagger}}
\right) \times
\nonumber \\
&& 
\left[\mathsf{n}^2 - (\mathsf{n}^{eq})^2\right]\,.
\label{BE_T:app}
\end{eqnarray}
The quantity $\beta(T) = \dfrac{g^{1/2}_{\star}(T)
\sqrt{g_{\rho}(T)}}{g_s(T)}$ where $g_s$ and $g_{\rho}$
are effective DOFs associated with entropy and energy
densities respectively and $g^{1/2}_{\star} =
\dfrac{g_s}{\sqrt{g_{\rho}}}\left(1+\dfrac{1}{3}
\dfrac{T}{g_s}\dfrac{dg_s}{dT}\right)$.
Finally, let us rewrite the above equation in terms of
previously defined dimensionless variables namely $x=M_{0}/T$,
$Y=\mathsf{n}/{\rm s}$ and $\xi = T_{\nu_R}/T$. In terms of
$x$, $Y$ and $\xi$ the Boltzmann equation can be expressed as
\begin{eqnarray}
x\,\dfrac{d\xi}{dx} + (\beta-1)\xi &=&
\dfrac{1}{2}\,\dfrac{\beta\,x^4\,{\rm s}^2}
{4\,\alpha\,\xi^3\,\mathcal{H}\,M_{0}^4}
\langle {E\sigma {\rm v}} \rangle_{eff}
\left[Y^2 - (Y^{eq})^2\right]\,.
\label{BE_xi}
\end{eqnarray}
Here, $\langle E \sigma {\rm v} \rangle_{eff}$ is define as
\begin{eqnarray}
\langle {E\sigma {\rm v}} \rangle_{eff} = 
\dfrac{(\mathsf{n}^{eq}_{\nu_R})^2}
{\left(\mathsf{n}^{eq}_{\phi} + 
\mathsf{n}^{eq}_{\psi}\right)^2}\times
\left(\langle{E\,\sigma{\rm v}}\rangle_{
\nu_R \overline{\nu_R}\rightarrow\psi\overline{\psi}}\,
+ 
\langle{E\,\sigma{\rm v}}\rangle_{\nu_R \overline{\nu_R}
\rightarrow \phi\phi^\dagger}\right)
\,.
\label{esigmaVeff:app}
\end{eqnarray}
The expression of $\langle {E\sigma {\rm v}} \rangle_{eff}$
can also be written in the following form suitable for numerical
computations as
\begin{eqnarray}
\langle {E\sigma {\rm v}} \rangle_{eff} = 
\left(\dfrac{\mathsf{n}_{\psi}^{eq}}
{\mathsf{n}_{\psi}^{eq} + \mathsf{n}_{\phi}^{eq}}\right)^2
\langle{E\,\sigma{\rm v}}\rangle^\prime_{
\nu_R \overline{\nu_R}\rightarrow\psi\overline{\psi}}\,
+ 
\left(\dfrac{\mathsf{n}_{\phi}^{eq}}
{\mathsf{n}_{\psi}^{eq} + \mathsf{n}_{\phi}^{eq}}\right)^2
\langle{E\,\sigma{\rm v}}\rangle^\prime_{\nu_R \overline{\nu_R}
\rightarrow \phi\phi^\dagger}
\,,
\label{esigmaVeff_numeric}
\end{eqnarray}
where, $\langle{E\,\sigma{\rm v}}\rangle^\prime_{
\nu_R \overline{\nu_R}\rightarrow j\overline{j}}$ is the
thermal average of $E \times {\sigma {\rm v}}_{\nu_R
\overline{\nu_R}\rightarrow j\overline{j}}$ normalised
by the product of equilibrium number densities of the final
state particles i.e. $n^{eq}_{j}n^{eq}_{\bar{j}}$.
\section{$2\rightarrow2$ scattering cross sections for
thermalisation of $\phi$}
\label{App:2to2scatt}
In this section we have listed expressions of all
$2\rightarrow2$ elastic scattering cross sections involving
$\phi$ to thermalise with the SM sector. The Feynmann
diagrams are shown in Fig.\,\,\ref{Fig:feyn_dia1}.  
\begin{eqnarray}
&& \hspace{-1.5cm}
\sigma_{\phi f \rightarrow \phi f} =
\frac{\lambda_{H\phi}^2 M_f^2}{16\pi}
\left[\frac{4 M_f^2- M_h^2}{M_h^2 
\left(M_f^4-2 M_f^2 \left(M^2_\phi+s\right)
+ M_h^2 s+\left(M^2_\phi-s\right)^2\right)}
\right. \nonumber \\ && ~~~~~~~~~~~~~~\left.
+\frac{\log \left(\frac{M_f^4-2 M_f^2
\left(M^2_\phi+s\right)+M_h^2 s+
\left(M^2_\phi-s\right)^2}{M_h^2 s}\right)}
{M_f^4-2 M_f^2\left(M^2_\phi+s\right)
+\left(M^2_\phi-s\right)^2}\right]\,,
\label{phif2phif}\\
&& \hspace{-1.5cm}
\sigma_{\phi V \rightarrow \phi V} =
\frac{\lambda_{H\phi}^2}{48 \pi}\frac{1}
{\left(\left({M_V}^2+{M_\phi}^2-s\right)^2
-4 {M_V}^2 {M_\phi}^2\right)} \times \nonumber \\
&&~~~~~~~~~\left[
\frac{\Lambda(s, M^2_V, M^2_{\phi})
\left \{ 2 {M_h}^4 s+{M_h}^2 
\left({M_V}^4-2 {M_V}^2 \left(\text{M$\phi $}^2+3 s\right)
+\left({M_{\phi}}^2-s\right)^2\right)+12 {M_V}^4 s\right \}}
{{M_h}^2 s \left({M_h}^2 s+{M_V}^4-2 {M_V}^2
\left({M_{\phi}}^2+s\right)+\left({M_{\phi}}^2-s\right)^2\right)}
\right. \nonumber \\ && \left. 
~~~~~~~~~   
-2 \left({M_h}^2-2 {M_V}^2\right) 
\log \left(\frac{{M_h}^2 s+ {M_V}^4-2 {M_V}^2 
\left({M_{\phi}}^2+s\right)+\left({M_{\phi}^2}-s\right)^2}
{{M_h}^2 s}\right)\right]\,, 
\label{phiV2phiV}\\
&& \hspace{-1.5cm}
\sigma_{\phi h \rightarrow \phi h} =
\frac{{\lambda_{H\phi}^2}}{16 \pi }
 \left[\frac{1}{s} -
 \frac{6 M_h^2 \log \left(\frac{M_h^4-M_h^2 
 \left(2 M^2_\phi+s\right)+\left(M^2_\phi-s\right)^2}
 {M_h^2 s}\right)}{M_h^4-2 M_h^2 \left(M^2_\phi+s\right)
 +\left(M^2_\phi-s\right)^2}
 + \frac{9 M_h^2}{M_h^4-M_h^2 \left(2 {M_{\phi}^2}+s\right)
 +\left({M^2_{\phi}}-s\right)^2} 
 \right]\,, \nonumber
 \label{phih2phih}\\
\end{eqnarray} 
where, $\Lambda(a,b,c)= (a -b -c)^2 - 4 bc$. The SM fermions
are gauge bosons are denoted by $f$ and $V$ respectively. In the
expressions of $\sigma_{\phi h \rightarrow \phi h}$ we have
considered terms up to $\mathcal{O}(\lambda^2_{H\phi})$ as in our
case the portal coupling $\lambda_{H\phi} << 1$.
\section{$2\rightarrow 2$ Annihilation cross sections
among $\mathbb{Z}_4$ charged particles}
\label{App:2to2xsections}
In this section we have listed the expressions
of all $2\rightarrow2$ annihilation cross sections
involving $\mathbb{Z}_4$ charged particles, which
are required for solving Eq.\,\,(\ref{BEy_T_lt_Tdec}) and
Eq.\,\,(\ref{BExi}). The Feynman diagrams are shown in Fig.\,\,\ref{Fig:feyn_dia1}.
\begin{eqnarray}
\sigma_{\phi\phi^\dagger\rightarrow \nu_R\overline{\nu_R}} =
\frac{y_{\phi}^4 \left(-\frac{\sqrt{s(s-4 M_{\phi}^2)}
   \left(M_{\psi}^2 s+2 \left(M_{\phi}^2-M_{\psi}^2\right)^2\right)}
   {M_{\psi}^2 s+\left(M_{\phi}^2-M_{\psi}^2\right)^2}
   +\left(2 M_{\psi}^2-2 M_{\phi}^2+s\right) \log
   \left(\frac{2 M_{\psi}^2+\sqrt{s(s-4 M_{\phi}^2)}-2
   M_{\phi}^2+s}{2 M_{\psi}^2-\sqrt{s(s-4 M_{\phi}^2)}-2
   M_{\phi}^2+s}\right)\right)}{8 \pi  s \left(s-4
   M_{\phi}^2\right)} \,\,,\nonumber 
   \label{phiphi2nuRnuRbar} \\
\end{eqnarray}
\begin{eqnarray}
\sigma_{\psi\overline{\psi}\rightarrow \nu_R\overline{\nu_R}} =
\frac{y_{\phi}^4 \left(\frac{\sqrt{s(s-4 M{\psi}^2)} 
\left(2\left(M_{\phi}^2-M_{\psi}^2\right)^2+M_{\phi}^2 s\right)}
{\left(M_{\phi}^2-M_{\psi}^2\right)^2+ M_{\phi}^2 s}
+2\left(M_{\psi}^2-M_{\phi}^2\right) 
\log \left(\frac{\sqrt{s(s-4 M_{\psi}^2)}-
2 M_{\psi}^2+2 M_{\phi}^2+s}{-\sqrt{s(s-4 M_{\psi}^2)}
-2 M_{\psi}^2+2 M_{\phi}^2+s}\right)\right)}
{16 \pi  s \left(s-4 M_{\psi}^2\right)} \,, \nonumber
\label{psipsibar2nuEnuRbar} \\
\end{eqnarray}
\begin{eqnarray}
\sigma_{\nu_R\overline{\nu_R}\rightarrow \phi\phi^\dagger} =
\frac{y_{\phi}^4 \left(-\frac{\sqrt{s(s-4 M_{\phi}^2)}
   \left(M_{\psi}^2 s+2 \left(M_{\phi}^2-M_{\psi}^2\right)^2\right)}
   {M_{\psi}^2 s+\left(M_{\phi}^2-M_{\psi}^2\right)^2}
   +\left(2 M_{\psi}^2-2 M_{\phi}^2+s\right) \log
   \left(\frac{2 M_{\psi}^2+\sqrt{s(s-4 M_{\phi}^2)}-2
   M_{\phi}^2+s}{2 M_{\psi}^2-\sqrt{s(s-4 M_{\phi}^2)}-2
   M_{\phi}^2+s}\right)\right)}{32 \pi  s^2} \,\,,\nonumber 
   \label{nuRnuRbar2phiphi} \\
\end{eqnarray}
\begin{eqnarray}
\sigma_{\nu_R\overline{\nu_R}} \rightarrow \psi \overline{\psi}=
\frac{y_{\phi}^4 \left(\frac{\sqrt{s(s-4 M_{\psi}^2)} 
\left(2\left(M_{\phi}^2-M_{\psi}^2\right)^2+M_{\phi}^2 s\right)}
{\left(M_{\phi}^2-M_{\psi}^2\right)^2+ M_{\phi}^2 s}+2
\left(M_{\psi}^2-M_{\phi}^2\right)
\log \left(\frac{\sqrt{s(s-4 M_{\psi}^2)}
-2 M_{\psi}^2+2 M_{\phi}^2+s}{-\sqrt{s(s-4 M_{\psi}^2)}
-2 M_{\psi}^2+2 M_{\phi}^2+s}\right)\right)}{16 \pi  s^2} \,.\nonumber \\
\label{nuEnuRbar2psipsibar}
\end{eqnarray}

\bibliographystyle{JHEP}
\bibliography{ref.bib} 
\end{document}